\documentclass{article}
\usepackage{arxiv}
\usepackage{amsmath}
\usepackage{dsfont}
\usepackage{array}
\usepackage{graphicx}
\usepackage{adjustbox}
\usepackage{tikz}
\usepackage{mathtools}
\usepackage[title]{appendix}
\usepackage{booktabs}
\usepackage{hyperref}
\usepackage[nameinlink]{cleveref}
\usepackage{natbib}
\usepackage{hyperref}
\usepackage[]{algorithm2e}
\makeatletter
\renewcommand{\@algocf@capt@plain}{above}
\makeatother
\usetikzlibrary{backgrounds,automata,arrows,positioning,calc}
\RestyleAlgo{boxruled}
\usepackage{subcaption} 
\usepackage[symbol]{footmisc}

\usetikzlibrary{backgrounds,automata,arrows,positioning,calc}

\DeclarePairedDelimiter\floor{\lfloor}{\rfloor}
\newcommand{\frap}[1]{\{#1\}}
\newcommand{\estat}[2]{\hat{S}^{#2}_{#1}}
\newcommand{\stat}[2]{S^{#2}_{#1}}
\newcommand{\ground}{{}_{\downarrow}\hspace{-0.5mm}}
\newcommand{\mint}[1]{\Lambda_{#1}}
\newcommand{\gmint}[1]{\Lambda_{\hspace{-0.5mm}\ground#1}}
\newcommand{\fmm}[2]{\alpha^{(#1)}_{#2} }
\newcommand{\sigfrac}{\eta}
\newcommand{\gfmm}[2]{\alpha^{(#1)}_{\hspace{-0.5mm}\ground#2} }
\newcommand{\eintf}[1]{\hat{\lambda}_{\hspace{-0.5mm}\ground#1}}
\newcommand{\intf}[1]{\lambda_{\hspace{-0.5mm}\ground#1}}
\newcommand{\epcf}[1]{\hat{g}_{\hspace{-0.5mm}\ground#1}}
\newcommand{\pcf}[1]{g_{\hspace{-0.5mm}\ground#1}}
\newcommand{\palm}[3]{M^{(#1)}_{#2|#3}}
\newcommand{\palms}[2]{M^{(#1)}_{#2}}
\newcommand{\emcor}[2]{\hat{k}_{#1}^{#2}}
\newcommand{\mcor}[2]{k_{#1}^{#2}}

\newcommand{\dispdense}{h_{\epsilon}}
\newcommand{\smean}[2]{\mathds{E}_{#1}\left[ #2  \right] }
\newcommand{\rmean}[1]{\mathds{E}\left[ #1  \right] }

\newcommand{\depmark}[1]{\gamma_1(#1)}
\newcommand{\indmark}[1]{\gamma_2(#1)}
\newcommand{\N}{\mathds{N}}
\newcommand{\Z}{\mathds{Z}}
\newcommand{\R}{\mathds{R}}

\title{Semiparametric point process modeling of blinking artifacts in PALM}
\author{
	Louis G. Jensen*\\
	Department of Mathematics \\
	Aarhus University \\
	Aarhus C \\
	Denmark \\
	\And
	David J. Williamson \\
	Randall Division for Cell and Molecular Biophysics\\ 
	King's College London\\ 
	London \\
	UK \\
	\And
	Ute Hahn \\
	Department of Mathematics \\
	Aarhus University \\
	Aarhus C \\
	Denmark \\
}

\begin{document}
\maketitle
\footnotetext{* Corresponding author \\ \textit{Email address:} \texttt{Louis.Gammelgaard@gmail.com}}
\begin{abstract}
    Photoactivated localization microscopy (PALM) is a powerful imaging technique for characterization of protein organization in biological cells. Due to the stochastic blinking of fluorescent probes, and camera discretization effects, each protein gives rise to a cluster of artificial observations. These blinking artifacts are an obstacle for quantitative analysis of PALM data, and tools for their correction are in high demand. We develop the Independent Blinking Cluster point process (IBCpp) family of models, which is suited for modeling of data from single-molecule localization microscopy modalities, and we present results on the mark correlation function. We then construct the PALM-IBCpp - a semiparametric IBCpp tailored for PALM data, and we describe a procedure for estimation of parameters, which can be used without parametric assumptions on the spatial organization of proteins. Our model is validated on nuclear pore complex reference data, where the ground truth was accurately recovered, and we demonstrate how the estimated blinking parameters can be used to perform a blinking corrected test for protein clustering in a cell expressing the adaptor protein LAT. Finally, we consider simulations with varying degrees of blinking and protein clustering to shed light on the expected performance in a range of realistic settings.
\end{abstract}
\noindent%

\section{Introduction}
Breaking the resolution limit imposed on classical fluorescence microscopy has been made possible by the advent of super resolution methods \citep{Huang2009}. Among these, PALM \citep{Betzig2006} has become a popular tool for the acquisition of point maps of individual molecules, achieved by the use of photoactivatable fluorescent proteins (PA-FPs). PA-FPs can be activated, read, and permanently photobleached in stochastic fashion. The resulting separation of fluorescent signal in time-space will, with high probability, be sufficient to individually localize the PA-FPs present in a given sample \citep{Yamanaka2014}. 

Unfortunately, it is the nature of PA-FPs to enter and reemerge from dark states a number of times before permanently bleaching, leading to multiple appearances of the same protein \citep{Annibale2011a, Fricke2015}. For analysis of the spatial organization of molecules, these reappearances lead to erroneous conclusions, unless explicitly dealt with \citep{Shivanandan2014}. In particular, analysis of the clustering properties of proteins, a common goal in PALM studies, is an increasingly contentious topic \citep{Rossboth2018}. Making matters worse, direct modeling of the blinking artifacts is complicated due to camera discretization of the continuous fluorescent signals \citep{Griffi_2020, Patel2019a}, and an understanding of both PA-FP photophysics and discretization effects is required to properly remedy the situation. 

Although such artifacts are best understood by considering the spatio-temporal behavior of PA-FPs, established methods for analysis of blinking artifacts have so far focused on one dimension or the other. In methods such as \citep{Andersen2018, Sengupta2011}, the spatial data alone is used, and require a model for protein behavior. Other methods use the temporal fluorescence traces to estimate the number of proteins in local regions \citep{Hummer2016, Karathanasis2017, Lin2015}, which require either manual segmentation or external calibration samples. More recently, complex descriptions of PA-FP photophysics have been modeled by means of Hidden Markov Models (HMM) \citep{Staudt2020, Patel2019a}. In \citep{Patel2019a}, estimation is carried out by means of a calibration sample of well-separated fluorophores. More recently, \citep{Staudt2020} model the conglomerate fluorescent intensity trace over a sequence of time points, as originating from some unknown number of PA-FP. This means that additional parameters have to be estimated, and the information in the spatial dimension is not exploited. 

In this paper, we define the family of Independent Blinking Cluster point processes (IBCpp) for single-molecule localization microscopy (SMLM) data, and present a result on the mark correlation function that is useful for estimation. We propose a particular model from the family, the PALM-IBCpp, for modeling of PALM data, and motivate the construction in terms of a discretized, 4-state PA-FP blinking model. We present an algorithm for estimating the parameters that control data artifacts, which can be run quickly even on large datasets. Our approach leads to estimates of the kinetic rates that govern photoblinking, which can be used to quantify the effect of blinking artifacts on a given sample, and correct downstream analyses for blinking induced biases. The modeling efforts are validate on established reference data of nuclear pore complexes (NUP) \citep{Thevathasan2019}. 

To help facilitate the debate on whether real protein clustering is present in a given sample, we devise a blinking corrected test for complete spatial randomness (CSR) on the basis of estimated blinking dynamics, and demonstrate it on a real biological sample of a cell expressing the protein Linker for Activation of T cells (LAT), observed at the plasma membrane. In this way, we can show that there are both areas of significant and non-significant protein clustering at different sites in the cell. This analysis serves as an example on the use of this universal test, and additionally provides yet more evidence for protein cluster in LAT, a research area of interest in its own right \citep{Williamson2011}. 

The paper is organized as follows. In \Cref{sec:prereqs}, we briefly go over the needed point process theory that will be used for modeling or estimation, and we give a quick rundown of the principles of PALM imaging, and how camera artifacts come into play. In \Cref{sec:model}, we define the IBCpp class of models, and present a useful result on the mark correlation function. We then construct and motivate the PALM-IBCpp for modeling of PALM data.. In \Cref{sec:est}, we describe an algorithm for estimation of the kinetic rates in the PA-FP blinking model. We validate our methods on nuclear pore complex reference data in \Cref{sec:NUP} by demonstrating a close alignment with expected blinking targets. \Cref{sec:data_sect} considers a dataset expressing LAT-mEos3.2 PA-FP, and we demonstrate how a blinking corrected CSR test can be performed on the basis of estimated blinking dynamics. Finally, in \Cref{sec:simstud}, we simulate PA-FP with a range of different spatial organizations and blinking behaviors, and illustrate the ability of our estimation methods to precisely recover the kinetic rates. We also consider what happens when the blinking model is misspecified, and we find that important PA-FP descriptors, such as the total number of reappearances and time to activation and bleaching, can still be recovered.

\section{Prerequisites} \label{sec:prereqs}
In this section we present the notation and point process concepts that we will be needing below, including moment measures, mark distributions, and the mark correlation function. We also describe some of the modeling difficulties that arise in SMLM experiments, namely those associated with discretization of the temporal information and background noise. For the general exposition, we work with processes on $\R^d\times \R_+$, but it is instructive to imagine $d = 2$, corresponding to 2D microscopy, which is the most common modality. For a more rigorous introduction to point process theory, we refer to \citep{Daley2007}. For more on mark distributions, see \citep{Stoyan1984}. Finally, more on the acquisition and preparation of SMLM data can be found in \citep{Deschout2014}. 

\subsection{Point processes and moment measures}
For the purpose of this paper, a spatio-temporal point process, $V = \{(v_i,t_{v_i})\}_{i = 1}^{\infty}$, is a random, locally finite point configuration with distinct points in $\R^d\times\R_+$. We call $V$ stationary if 
\begin{equation}
V \stackrel{d}{=} V+s = \{v_i+s, t_{v_i}\}_{i = 1}^{\infty}, 
\end{equation}
for all $s \in \R^d$, where $\stackrel{d}{=}$ denotes equality in distribution. Similarly, we call $V$ rotation-invariant if 
\begin{equation}
V \stackrel{d}{=} RV = \{Rv_i, t_{v_i}\}_{i = 1}^{\infty},
\end{equation}
for any rotation $R$. If $V$ is both stationary and rotation-invariant, it is motion-invariant.

Write $\ground V = \{v_i\}_{i = 1}^{\infty}$ (\textit{ground V}) for the random object obtained by stripping $V$ of its times. Assume $\ground V$ is well-defined as a spatial point process on $\R^d$, having finite intensity function $\intf{V}$ and second-order product density $\intf{V}^{(2)}$. Then we compute the (ground) intensity measure, $\gmint{V}$, and (ground) second-order factorial moment measure, $\gfmm{2}{V}$, as
\begin{align}
\gmint{V}(A) &=\rmean{\sum_{v \in \ground V} \mathds{1}_A(v)} =  \int_A \intf{V}(v) dv,\\
\gfmm{2}{V}(A_1\times A_2)& = 
\rmean{\sum_{(v_1,v_2) \in \ground V^2}^{\neq} \mathds{1}_{A_1\times A_2}(v_1,v_2)}
=\int_{A_1\times A_2} \intf{V}^{(2)}(v_1,v_2)d(v_1,v_2),
\end{align}
working everywhere on Borel sets, and $\sum^{\neq}$ means summation over distinct pairs of points. The pair correlation function $\pcf{V}$ is then defined in the usual way
\begin{equation}
\pcf{V}(v_1,v_2) = \frac{\intf{V}^{(2)}(v_1,v_2)}{\intf{V}(v_1)\intf{V}(v_2)}.
\end{equation}
Next, the 1-point mark distribution, $\palm{1}{V}{v}$, is defined via the space-time intensity measure. When it exists, it is the conditional probability measure on $\R_+$ satisfying
\begin{equation}
\mint{V}(A\times B) = \rmean{\sum_{(v,t_v)\in V} \mathds{1}_A(v)\mathds{1}_B(t_v)} =\int_A \palm{1}{V}{v}(B) d \gmint{V}(v)	.
\end{equation}
Similarly, the \textit{2-point mark distribution}, $\palm{2}{V}{(v_1,v_2)}$, satisfies the conditional measure representation of the space-time second-order factorial moment measure
\begin{align} \label{2palmmark}
\fmm{2}{V}(\times _{k = 1}^2[A_k\times B_k]) &= 
\rmean{\sum_{(v_1,t_{v_1}),(v_2,t_{v_2}) \in V^2}^{\neq} \mathds{1}_{A_1\times A_2}(v_1,v_2)\mathds{1}_{B_1\times B_2}(t_{v_1},t_{v_2})}
\\ &= \int_{A_1\times A_2} \palm{2}{V}{(v_1,v_2)}(B_1\times B_2) d\gfmm{2}{V}(v_1,v_2).
\end{align}
From these conditional measures, the mark correlation function, $\mcor{V}{f}$, is defined as 
\begin{equation}
\mcor{V}{f}(v_1,v_2) = \frac{\int f(t_{v_1},t_{v_2}) d\palm{2}{V}{(v_1,v_2)}(t_{v_1},t_{v_2})}{\int\int f(t_{v_1},t_{v_2}) d\palm{1}{V}{v_1}(t_{v_1})d\palm{1}{V}{v_2}(t_{v_2})},
\end{equation}
for $f : \R_+^2 \mapsto \R_+$ a non-negative Borel function of two times. We will refer to $f$ as a \textsl{query function}.  


\subsection{PALM, discretization, and noise} \label{art_gritty_sect}
To understand how PALM works, we imagine a single PA-FP located at the position $x$. Whenever fluorescence is emitted, it is captured by the camera, and the signal is integrated over the acquisition time lasting 1 frame. Based on the intensity profile observed on pixels, the position $x$ is estimated, by assuming a shape for the point spread function (PSF) \citep{Small2014, Ober_2015}, which models the blurry shape observed on a camera when imaging a point-source of light. The localization uncertainty associated with the estimate of $x$ can then be computed, and is included in the dataset for each localization. This localization procedure is possible because we assumed only a single, isolated fluorescent emitter. In a real biological sample, there can be several emitters at nearly the same position, and the assumption of an isolated signal is thus often violated. However, if we only receive a signal of finite length from each emitter, in non-overlapping windows of time, the spatial proximity becomes irrelevant, and we can again determine the position of each emitter. In PALM, this temporal separation is made possible using PA-FPs, which activate at different times, and turn off permanently after finite emission of fluorescence. In this way, only a single emitter should be active at a given space-time location, and it can then be precisely localized.

Note that, using the procedure outlined above, each emitter will give rise to several localizations. To see why this is true, assume that the PA-FP at position $x$ sends out a (sufficiently bright) signal lasting in total $T$ seconds, and the frame acquisition time is $\Delta$ seconds. We can then expect the signal to result in roughly $T \Delta^{-1}$ estimates of $x$, all of which will be included in the sample as separate localizations. Depending on the total fluorescence observed from the PA-FPs, and the camera framerate, this can lead to a large number of reappearances per protein. It is natural to think that this problem can be solved by grouping localization that are close in space-time, and although such procedures are often used in practice \citep{Annibale2011, Lee2012}, they are typically heuristic in nature due to the lack of precise knowledge about the temporal behavior of the PA-FPs in the sample. Without such knowledge, we have no principled guide for determining the merging thresholds, which must allow both for varying spatial uncertainty, and extended temporal separation occurring due to PA-FPs visiting dark states. As a result, localizations arising from the same emitter can be easily confused with those arising from a nearby, or nonexistent, emitter. 

In addition to reappearances, background noise will invariably affect the dataset. Each time fluorescence is observed on the camera, it must be attributed as spurious background or coming from a PA-FP emission event, by means of a separating threshold. Since we cannot set the threshold too high without losing the signal of real PA-FP, some background noise points will always be present in PALM recordings. 

\section{Independent Blinking Cluster point processes}\label{sec:model}
In this section we introduce and motivate the IBCpp family of models, which is a subset of clustered spatio-temporal point processes with a particular spatio-temporal clustering structure that is natural for modeling of SMLM data. We then consider a moment result with particular importance for parameter estimation. Finally, we construct the PALM-IBCpp, which is a semiparametric IBCpp model tailored for PALM data.

\subsection{Definition}
A point process following the IBCpp model, denoted by $O$ throughout, has $3$ components: the process of protein locations, $\ground X$, the blinking cluster of all localizations and timepoints associated with a protein $x$, $Y_x$, and an independent Poisson process of noise points, $E$. The IBCpp $O$ is then constructed hierarchically as the union of all blinking clusters, $Z$, with the noise process, $E$, as
\begin{align}
O &=  Z \bigcup E, \\
Z &= \bigcup_{x \in X} Y_{x},
\end{align} 
where we assume the blinking clusters are independent of each other, and of the form
\begin{equation}
Y_{x} = \bigcup_{i = 1}^G (x+\epsilon_i, t_{y_i}),
\end{equation}
where the $\epsilon_i$ are i.i.d. with distribution $P_{\epsilon}$, and further independent of $\{t_{y_i}\}_{i = 1}^G$ and $G$. Finally, the spatio-temporal intensity of the noise process is assumed to be on the form
\begin{equation}
\lambda_E(e,t_e) = \intf{E}\frac{\mathds{1}(t_e \le b)}{b},
\end{equation}
where $b$ is the length (in seconds) of the data recording and $0 \le \intf{E} < \infty$. 

To explain why this construction is natural for SMLM data, we now consider each component and assumption above in more detail. Starting with the overall structure of $O$, essentially all SMLM modalities should be modeled naturally with this general idea of (possibly repeated) noisy observations of the proteins in the sample, corrupted by spurious background noise. This is certainly the case for commonly used modalities such as PALM, STORM \citep{Rust2006}, DNA-PAINT \citep{Schnitzbauer2017}, and many others. 

The real meat of the definition is in the parametrization of a blinking cluster, $Y_x$, and the dependence assumptions within and between different blinking clusters. Starting from the assumption of independently blinking fluorophores (and thus blinking clusters), this is a standard convenience assumption in the literature \citep{Rollins2015, Staudt2020}, albeit likely an approximation in samples with extreme local density. For the timepoints and the number of points in $Y_x$, $|Y_x| = G$, we allow general distribution and dependence structure. We need this level of generality as both are typically derived from the same, underlying source of stochasticity. Taking PALM as an example, the PA-FP in the sample switch between fluorescent and non-fluorescent states according to a continuous time absorbing Markov process, $S(t)$, and the observed times then correspond to the camera frames that overlap a fluorescent state visit. More broadly we can  imagine the observed timepoints in $Y_x$ arising as
\begin{equation}
D(S) = \{t_{y_1},t_{y_2},...,t_{y_G}\},
\end{equation}
where $D$ is a ''discretization operator'' (the camera, localization software, filtering,...), transforming $S$ into the observed signal. In particular, the distribution and dependence structures of $G$ and $\{t_{y_i}\}_{i = 1}^G$ are both derived in some complex way from the same stochastic process, see \Cref{fig:statediag} and \Cref{fig:discretesignal}.

Finally, for the locations in $Y_x$, $\{y_i\}_{i = 1}^G$, recall that positions are estimated on the basis of fitting to a blurry point spread function (PSF) centered on $x$. This motivates why the locations in $Y_x$ are modeled on the form
\begin{equation} \label{clusterlocs}
y_k = x+\epsilon_k,
\end{equation}
where $\epsilon_k$ is a random variable on $\R^d$ reflecting our uncertainty about the true position $x$. The distributional shape and scale of $\epsilon_k$ depends on the PSF and on the number of photons detected by the camera during the associated camera frame. As a practically necessary assumption, we modeled the collection $\{\epsilon_k\}_{k = 1}^G$ as i.i.d., and further independent of the timepoints and $G$. These assumptions can all be motivated by the time-homogeneous Markov processes underlying photon statistics \citep{Staudt2020}, which imply that the number of photons hitting different frames are approximately independent, and further independent of which frame number is currently being imaged.

\subsection{A result on the mark correlation function} \label{sec:moment}
Let O be an IBCpp with motion-invariant X. We present here a key result on the mark correlation function, which we use to motivate the estimation procedures of \Cref{sec:est}. The derivations of the results in this section and more can be found in \hyperref[suppsec]{Section A} of the supplementary material. 

Let $f: \R_+^2 \mapsto \R_+$ be a symmetric query function of 2 arrival times, and assume $P_{\epsilon}$ has radially symmetric density function $\dispdense$. Then, the pair- and mark correlation functions are functions only of the distance between two points, $r$, and for the product between them we have the result
\begin{equation} \label{IBCpp:gkg1}
\gamma_2^O(f)\mcor{O}{f}(r)\pcf{O}(r) = (\depmark{f}-\indmark{f})\left[\frac{\sigfrac}{\intf{O}}n_c(\dispdense*\dispdense)(r)\right] +\indmark{f}\left[\pcf{O}(r)-1\right]+\gamma_2^O(f)
\end{equation}
where 
\begin{align}
n_c &= \frac{\rmean{G^2}}{\rmean{G}}-1,\\
\sigfrac &= \frac{\intf{Z}}{\intf{O}}, \\
\depmark{f}  \label{depmark}
&= \frac{\rmean{\sum_{(i,j) = 1}^{G}\mathds{1}(i\neq j)f(t_{y_i}, t_{y_j})}}{\rmean{G(G-1)}}, \\
\indmark{f} \label{indmark}
&= \frac{\rmean{\sum_{i = 1}^{G}\sum_{i = 1}^{G'}f(t_{y_i}, t'_{y_j})}}{\rmean{G}^2}, \\
\gamma_2^O(f) &= \int\int f(t_{1},t_{2}) d\palms{1}{O}(t_{1})d\palms{1}{O}(t_{2}),	
\end{align}
and
\begin{equation} \label{locdens}
(\dispdense*\dispdense)(r) = \int \dispdense(y_1-x)\dispdense(y_2-x) dx,
\end{equation}
for $||y_1-y_2|| = r$. In the above, $(G, \{t_{y_i}\}_{i = 1}^{G})$ should be thought of as the timepoints in a typical blinking cluster $Y_x$ at arbitrary location $x$, and $(G', \{t'_{y_j}\}_{j = 1}^{G'})$ is an independent copy of $(G, \{t_{y_i}\}_{i = 1}^{G})$. Finally, $\palms{1}{O}$ is the 1-point mark distribution of $O$, which does not depend on the conditioning point, which is therefore omitted in the notation.

We unpack this result now in some detail, providing first some intuition on the involved quantities. We also cover some related moment expression that will be needed in the following. Starting with $\eta$, it is the expected fraction of points in $O$ that arose from blinking clusters (as opposed to background noise), and in particular we have the alternative expression
\begin{equation} \label{alteta}
\eta = 1-\frac{\intf{E}}{\intf{O}},
\end{equation}
as $1$ minus the expected fraction of noise points. This is a simple consequence of the fact that the points in $O$ are either from $Z$ or $E$, so that
\begin{equation} \label{mom1}
\intf{O} = \intf{Z}+\intf{E}.
\end{equation}
A useful related expression is
\begin{equation} \label{mom2}
\intf{Z} = \rmean{G}\intf{X},
\end{equation}
which states the natural result that the number of points (per area) from blinking clusters can be written as the number of proteins (per area) times the number of repeats per protein. 

Moving on to the second-order quantities, $\gamma_1(f)$ is essentially the mean value of $f(t_{y_1}, t_{y_2})$ when $(t_{y_1}, t_{y_2})$ are sampled randomly from the distinct pairs of timepoints in a typical blinking cluster. It should be clear that, depending on the choice of $f$, $\gamma_1(f)$ will contain information about the blinking dynamics of the fluorophores in the sample, a fact we will exploit for estimation. Similarly, $\gamma_2(f)$ is the mean value of $f(t_{y_1}, t'_{y_2})$ when the timepoints are sampled randomly from $2$ different (and thus independent) blinking clusters. Lastly, $\gamma_2^O(f)$ is as before, but where each timepoint is an independently sampled timepoint among all timepoints in $O$, including those from noise points - it is also known as the normalization constant of the mark correlation function. Lastly, the spatial term $(\dispdense*\dispdense)(r)$ is simply the autoconvolution of the localization uncertainty density. 

The expression in \Cref{IBCpp:gkg1} is important from the standpoint of semiparametric estimation due to the split of terms into products of spatial and temporal components. The temporal components (the $\gamma$'s) and the spatial components ($\pcf{O}$ and $(\dispdense*\dispdense)$) are in this sense separable, which hints at the possibility of extracting information about the temporal behavior of fluorophores, independently of their spatial coordinates. To make more explicit how this should be done, note the simple algebraic manipulation
\begin{equation} \label{IBCpp::eta}
(\gamma_1(f)-\gamma_2(f))n_c =
\frac{\left(\gamma_2^O(f)\mcor{O}{f}(r)\pcf{O}(r)-\indmark{f}\left[\pcf{O}(r)-1\right]-\gamma_2^O(f)\right)\intf{O}}{(\dispdense*\dispdense)(r)\sigfrac}.
\end{equation}
The significance of this identity is that the left hand side depends \textsl{only} on the process that generated blinking, whereas the right hand side can be estimated from $O$, without a need to model $\ground X$. The idea is then to set these estimated quantities, for various $f$, in relation to their theoretical value under the parameters of a specified blinking model. We show how to do this in more detail in \Cref{sec:est}.

\subsection{An IBCpp model for PALM data} \label{sec:timemodel}
In order to use the IBCpp family in practice, we get more specific about the construction of the blinking clusters. The choices we make here are based on realistic models for PALM fluorophore photophysics, camera discretization effects, and localization errors, and lead to the PALM-IBCpp model. The PALM-IBCpp is most appropriate for modeling of 2D data, as 3D PALM generally has unequal uncertainty in the $xy$ versus $z$ plane \citep{Shtengel2009}, and a radial noise profile is then no longer a valid assumption. However, so long as the noise profile in the $xy$ plane has no preferred direction on average, 3D data can be used without complication by simply discarding the $z$-coordinates.

As in the general IBCpp formulation, we write the typical blinking cluster on the form
\begin{equation}
Y_x = \bigcup_{k = 1}^G (x+\epsilon_k, t_{y_k}),
\end{equation}
and we need to specify the distributions of $\epsilon_k$, $G$, and $t_{y_k}$. Starting with $\epsilon_k$, recall that a point source of light appears as a blurry spot on the camera, with shape described by the PSF. For PALM data we model this PSF using a symmetric Gaussian with random variance $\sigma^2$. We model $\sigma$ as random since its magnitude depends on the number of photons detected and various other nuisance factors that will vary for each observation. Denoting by $P_{\sigma}$ the distribution of $\sigma$, we thus write
\begin{align}
(\epsilon_k|\sigma) &\sim \mathrm{N}(0, \sigma^2), \\ 
\sigma &\sim P_{\sigma},
\end{align}
where $(\cdot|\sigma)$ denotes the $\sigma$-conditional distribution, and $\mathrm{N}(0,\sigma^2)$ is the centered Gaussian distribution with variance $\sigma^2$. 
Since localization software outputs an estimate $\hat{\sigma}$ for each observation, we do not need to parametrize $P_{\sigma}$. The use of Gaussian PSFs is standard practice, and generally provides a highly accurate approximation \citep{Zhang2007}, but another model for the PSF can be used without serious complications, so long as it is radially symmetric on average (across the typical observation).

\begin{figure}
	\centering
	\begin{minipage}[t]{1\textwidth}
		\centering
		\begin{tikzpicture}[->, >=stealth', auto, semithick, node distance=2cm]
		\tikzstyle{every state}=[fill=white,draw=black,thick,text=black,scale=1]
		\node[state]    (I)                {$I$};
		\node[state]    (F)[right = of I]  {\textcolor{red}{$F$}};
		\node[state]    (B)[right = of F]  {$B$};
		\node[state]    [above = 0.75cm of F] (D)  {$D$};
		\path
		(I) edge[]     node{$r_F$}     (F)
		(F) edge[bend right]     node{\hspace{-2cm}$r_{R}$}     (D)
		(D) edge[bend right]     node{\hspace{0.75cm}$r_{D}$}     (F)
		(F) edge[]     node{$r_B$}     (B);
		\end{tikzpicture}		
		\caption{The transition diagram for the continuous time, photophysical model of fluorophores. Transitions are Markovian, with rates indicated next to the transition arrows.}
		\label{fig:statediag}
	\end{minipage}
	\begin{minipage}[t]{1\textwidth}
		\centering
		\includegraphics[width=1\linewidth]{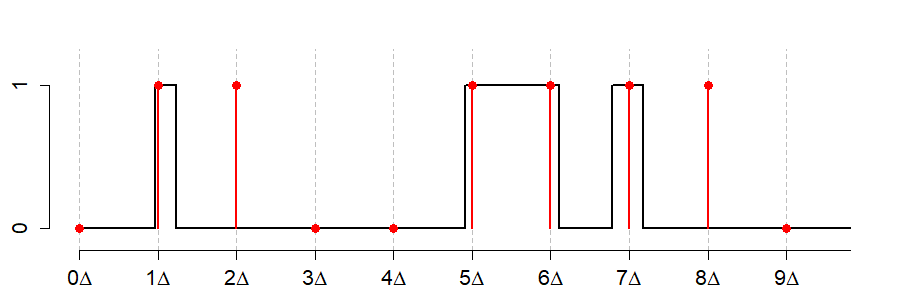}
		\caption{Camera discretization transforms the continuous process $S(t)$ (in black) into the discrete process $\tilde{S}(k\Delta)$ (in red). The observed timepoints are the $k\Delta$ with $\tilde{S}(k\Delta) = 1$; in this example there are $6$ such timepoints, observed on frames $\{1,2,5,6,7,8\}$, and we thus have $G = 6$ and $\{y_{t_k}\}_{k = 1}^{6} = (1\Delta,2\Delta,5\Delta,6\Delta,7\Delta,8\Delta)$. }
		\label{fig:discretesignal}
	\end{minipage}
\end{figure}

Moving on to $G$ and the timepoints, we take as basis a well-established 4-state model for continuous time fluorophore behavior \citep{Griffi_2020, Rollins2015, Coltharp2012}. We imagine the PA-FP are independently following a Markov processes, with a single fluorescent state $F$, and 3 non-fluorescent states, see \Cref{fig:statediag}. A PA-FP always begins in the inactive state $I$, and eventually moves to the $F$ state. From here, it can either go dark in $D$ temporarily, or permanently photobleach in $B$.

We cannot observe the process in continuous time. In fact, if we write $\Delta$ for the length of 1 camera frame, the temporal resolution allows observations to occur only on the fixed grid $\Delta\N$. To describe the fluorescent signal that is ultimately observed on this grid, from a single PA-FP during the experiment, we consider a discretization operation under an idealized camera. Consider the indicator process
\begin{equation}
S(t) = \begin{cases} 
1 & \text{if the PA-FP is in state $F$ at time $t$} \\
0                                    & \text{otherwise}.      %
\end{cases}
\end{equation}
We imagine that any (measurable) amount of fluorescent signal hitting a given camera frame gives rise to an observation. Defining
\begin{equation}
\tilde{S}(k\Delta) = \mathds{1}_{(0,\Delta]}\left( \int_{\Delta (k-1)}^{\Delta k} S(t) dt\right),
\end{equation}
the observed timepoints are then $k \Delta$ whenever $\tilde{S}(k\Delta) = 1$. This corresponds to a camera with perfect sensitivity, which is of course an approximation to the truth. In reality there is a non-zero threshold on the amount of signal that must be observed during a given integration period, but this threshold is generally very low in SMLM recordings \citep{Patel2019a}, so we have ignored it here to avoid the complications that arise from modeling it.

From the above, we can write $G$ and $\{t_{y_k}\}_{k = 1}^G$ more formally as
\begin{align}
G &= \sum_{k =1}^{\infty} \tilde{S}(k\Delta), \\
t_{y_k} &= \min \{s\Delta : s\in\N \text{ and } \sum_{i = 1}^{s}\tilde{S}(i\Delta) = k\}, \ 1 \le k \le G.
\end{align}
In this way, the timepoints of a typical cluster correspond precisely to the discretized signal obtained from $S(t)$, see \Cref{fig:discretesignal}.

\section{Estimation} \label{sec:est}
We suggest now a stepwise estimation procedure, leading eventually to estimates of $\eta$ and $(r_F, r_D, r_R, r_B)$. As the implementation details are somewhat long-winded, we describe the methods here at the intuitive level, and refer to supplementary \hyperref[suppsec]{Section B} and \hyperref[algofit]{Algorithm 1} for more details. For clarity of exposition, we motivate our approach on the assumption that $\ground X$ be motion-invariant, but we stress that this is not a necessary assumption in practice, as \hyperref[algofit]{Algorithm 1} will produce meaningful estimates also for general $\ground X$, cf. supplementary \hyperref[suppsec]{Section D}. Further, since the PALM-IBCpp is most appropriate for 2D data, as previously noted, we assume the spatial dimension is $d = 2$ in the following. An efficient implementation of \hyperref[algofit]{Algorithm 1}, and various other helpful tools, are available, see \hyperref[code]{R implementation}. 

\subsection{Data format and requirements}
In the following, we assume that we have data $\{(o_k, t_{o_k})\}_{i = 1}^N$ from a PALM-IBCpp observed with $N$ points in the space-time window $W\times [0,b]$, where $W\subset \R^2$ and $b$ is the length of the PALM recording in seconds. Additionally, we require access to localization uncertainties associated with each position, and we denote these by $\{\hat{\sigma}_k\}_{k = 1}^N$. Note that it is assumed the timepoints $t_{o_k}$ are recorded in seconds. Often it is the case that PALM data is recorded in terms of frame numbers, and it is then necessary to first transform the times by multiplying the frame numbers by the camera integration length, $\Delta$, which can be obtained from the framerate by
\begin{equation}
\Delta = \frac{1}{\text{framerate}},
\end{equation}
and is also a required component in its own right.

If the fitting procedures should account for background noise, it is also necessary to have access to an observation of pure noise, which will allow us to quantify the fraction of points arising as noise. Thus, we assume that we have $N_e$ observations $\{e_k,t_{e_k}\}_{k = 1}^{N_E}$ of $E$ in a separate space-time window $W_E \times [0,b]$. Access to $E$ in this way is typically possible without a need to perform additional experiments, as standard PALM recordings generally extend to regions outside the cell being imaged, see \Cref{fig:NUP} and \Cref{fig:dataset}. 
\subsection{Choice of query functions}\label{sec:sumstat}
The foundation for estimation of kinetic rates is the identity in \Cref{IBCpp::eta}, which allow us to extract a purely temporal information from the observed space-time data, principally via $\gamma_1(f)$ and $n_c$. The type and quality of this information depends crucially on our choice for the query function $f$. In the following, we pick the set of functions
\begin{align}
f_u(t_1,t_2) &= \mathds{1}(|t_1-t_2| \le u), \quad u \in T, \\
T &= \{i\Delta\}_{i = 1}^{\floor{\frac{b}{\Delta}}}.
\end{align}
This choice exhausts the information present in functions acting on times only through their difference, while eliminating absolute time information. To see why this can be desirable, imagine a typical blinking cluster $Y_x$. The timepoints in $Y_x$ can be written approximately (up to rounding-induced errors) on the form
\begin{equation}
t_{o_k} \approx W_F+w_k,
\end{equation}
\begin{algorithm}[H] \label{algofit}
	\small
	\SetKwBlock{kwInit}{Initialization}{end}
	\SetKwBlock{kwNoise}{Estimation of $\boldsymbol{\eta}$}{end}
	\SetKwBlock{kwRates}{Estimation of kinetic rates}{end}
	\SetKwInput{kwOutput}{Output}
	\caption{PALM-IBCpp model fit, part 1}
	\SetKwInOut{Input}{Input}
	\SetKwInOut{Output}{Output}
	\Input{Space-time observations $\{o_k, t_{o_k}\}_{k = 1}^N$ observed in window $W\times [0,b]$, where $b$ is the length of the PALM recording in seconds.} 
	\Input{Localization uncertainties $\{\hat{\sigma}_k\}_{k = 1}^N$.}
	\Input{Camera integration length, $\Delta = \frac{1}{\text{framerate}}$.}
	\Input{Quality parameters $n_r$ and $n_{s}$ (default values of $500$ and $10000$ are used everywhere in this work, respectively).}
	\Input{(optional) The noise process $E = \{e_k, t_{e_k}\}_{k = 1}^{N_E}$ observed separately in window $W_E \times [0,b]$.}
	\Output{Estimated fraction of non-noise points $\hat{\eta}$ and kinetic rates $(\hat{r}_F, \hat{r}_D, \hat{r}_R, \hat{r}_B)$.}
	\kwInit{
		\textbf{(1)} If $\{t_{o_k}\}_{k= 1}^N$ are stored as frame numbers, update each timepoint as
		\begin{equation*}
		t_{o_k} \leftarrow t_{o_k}\Delta.
		\end{equation*}\\
		\noindent \textbf{(2)} Define the spatial range $r_{\mathrm{max}}$ and grid $R$, by
		\begin{align*}
		&r_{\mathrm{max}}  = \frac{1}{N} \sum_{k = 1}^{N} \hat{\sigma}_k, 
		&R        = \{\frac{r_{\mathrm{max}}}{n_r} i\}_{i = 1}^{n_r}.
		\end{align*} \\
		\noindent \textbf{(3)} Define the temporal grid $T$ and query functions $f_u$ for $u\in T$ by
		\begin{align*}
		T &= \{\Delta i\}_{i = 1}^{\floor{\frac{b}{\Delta}}},
		&f_u(t_1,t_2) = \mathds{1}(|t_1-t_2| \le u).
		\end{align*}
	}
	\kwNoise{
		Set $\eintf{O} = \frac{N}{|W|}$. If $E$ was observed in a separate window, set $\eintf{E} = \frac{N_E}{|W_E|}$, and otherwise set $\eintf{E} = 0$. Return the estimator
		\begin{equation*}
		\hat{\eta} = 1-\frac{\eintf{E}}{\eintf{O}}.
		\end{equation*}
	}
\end{algorithm}
\noindent where $W_F \sim \mathrm{Exp}(r_F)$ is the time spent in the inactive $I$ state before first activation, and $w_k$ is the waiting time separating the $k$'th appearance from the temporal origin, which depends only on the remaining rates $(r_D,r_R,r_B)$. When extracting information from a query function through $\gamma_1(f)$, we then obtain
\begin{equation}
\gamma_1(f) \approx \frac{\rmean{\sum_{(i,j) = 1}^{G}\mathds{1}(i\neq j)f(W_F+w_i, W_F+w_j)}}{\rmean{G(G-1)}}.
\end{equation}
Since $r_F$ is typically orders of magnitudes smaller than the remaining rates, $W_F$ will tend to dominate and obscure the information on the remaining parameters. On the other hand, for $f_u$ we have
\begin{equation}
\gamma_1(f_u) \approx \frac{\rmean{\sum_{(i,j) = 1}^{G}\mathds{1}(i\neq j)\mathds{1}(|w_i-w_j| \le u)}}{\rmean{G(G-1)}},
\end{equation}

\setcounter{algocf}{0}
\begin{algorithm}[H] \label{algofit2}
	\small
	\SetKwBlock{kwInit}{Initialization}{end}
	\SetKwBlock{kwNoise}{Estimation of $\boldsymbol{\eta}$}{end}
	\SetKwBlock{kwRates}{Estimation of kinetic rates}{end}
	\SetKwInput{kwOutput}{Output}
	\caption{PALM-IBCpp model fit, part 2}
	\SetKwInOut{Input}{Input}
	\SetKwInOut{Output}{Output}
	\kwRates{
		\textbf{(1)} Let $\{\hat{\sigma}_{1,k}\}_{k = 1}^{n_{s}}$ and $\{\hat{\sigma}_{2,k}\}_{k = 1}^{n_{s}}$ be independent samples of size $n_{s}$ with replacement from $\{\hat{\sigma}_k\}_{k = 1}^N$. Estimate the blinking cluster autoconvolution via
		\begin{equation*}
		\widehat{(\dispdense\ast\dispdense)}(r) = \frac{1}{n_{s}}\sum_{k = 1}^{n_{s}} \frac{e^{-\frac{r^2}{2(\hat{\sigma}_{1,k}^2+\hat{\sigma}_{2,k}^2)}}}{2\pi (\hat{\sigma}_{1,k}^2+\hat{\sigma}_{2,k}^2)}, \quad r\in R.
		\end{equation*}\\
		\textbf{(2)} Let $\{t_{o_{1,k}}\}_{k = 1}^{n_{s}}$ and $\{t_{o_{2,k}}\}_{k = 1}^{n_{s}}$ be independent samples of size $n_{s}$ with replacement from $\{t_{o_k}\}_{k = 1}^N$. Estimate $\gamma_2^O(f_u)$ via
		\begin{equation*}
		\hat{\gamma}_2^O(f_u) = \frac{1}{n_{s}}\sum_{k = 1}^{n_{s}} \mathds{1}(|t_{1,k}-t_{2,k}| \le u), \quad u\in T.
		\end{equation*}
		\\
		\textbf{(3)} Define the distribution function
		\begin{equation*}
		\hat{M}_Z^{(1)}(u) = \frac{N^{-1}\sum_{k = 1}^{N}\mathds{1}(t_{o_k} \le u)-(1-\hat{\sigfrac})\frac{u}{b}}{\hat{\sigfrac}}.
		\end{equation*}
		Sample i.i.d. collections of variates $\{\tilde{t}_{1,k}\}_{k = 1}^{n_s}$ and $\{\tilde{t}_{2,k}\}_{k = 1}^{n_s}$ with distribution $\hat{M}_Z^{(1)}$, and use the estimator
		\begin{equation*}
		\hat{\gamma}_2(f_u) = \frac{1}{n_{s}}\sum_{k = 1}^{n_{s}} \mathds{1}(|\tilde{t}_{1,k}-\tilde{t}_{2,k}| \le u), \quad u\in T.
		\end{equation*}\\
		\textbf{(4)} Using (any) standard estimators for the mark- and pair correlation functions, $\emcor{O}{f}$ and $\epcf{O}$, set 
		\begin{equation*}
		\hat{\zeta}_u = \frac{\eintf{O}}{\hat{\sigfrac}}\frac{\sum_{r \in R}\left[\hat{\gamma}_2^O(f_u)\emcor{O}{f_u}(r)\epcf{O}(r)-\hat{\gamma}_2(f_u)(\epcf{O}(r)-1)-\hat{\gamma}_2^O(f_u)\right]\left[\widehat{(\dispdense*\dispdense)}(r)\right]}{\sum_{r \in R} \left[\widehat{(\dispdense*\dispdense)}(r)\right]^2},
		\end{equation*}
		for each $u \in T$.
		\\
		\textbf{(5)}
		Using the approximate expressions for $\gamma_1(f_u)$ and $n_c$ in supplementary \hyperref[suppsec]{Section B}, solve the weighted least squares problem
		\begin{equation*}
		\min_{\hat{r}_D,\hat{r}_R,\hat{r}_B} \sum_{u \in T}\sum_{r \in R} \left(\frac{\hat{\zeta}_u}{\hat{\gamma_2(f_u)}}\right)^2\left(\hat{\zeta}_u-(\gamma_1(f_u)-\hat{\gamma}_2(f_u))n_c\right)^2,
		\end{equation*}
		to obtain estimators $(\hat{r}_D,\hat{r}_R,\hat{r}_B)$. 
		\\
		\textbf{(6)}
		Obtain an estimator of $r_F$ by setting
		\begin{equation*}
		\hat{r}_F = \left(\frac{\frac{1}{N}\sum_{k = 1}^{N}t_{o_k}-(1-\hat{\eta})\frac{b}{2} }{\hat{\eta}} -\hat{A}_2-\hat{B}_2\right)^{-1}.
		\end{equation*}
		where $\hat{A}_2$ and $\hat{B}_2$ are defined in supplementary \hyperref[suppsec]{Section B}.
		\\
		\textbf{(7)}
		Obtained a censoring-corrected estimate of $r_F$ by numerically solving 
		\begin{equation*}
		\frac{e^{\hat{r}^c_F b}-\hat{r}^c_F b-1}{\hat{r}^c_F(e^{\hat{r}^c_Fb}-1)}-\frac{1}{\hat{r}_F} = 0,
		\end{equation*}
		in $\hat{r}^c_F$ over the interval $(0, \hat{r}_F]$. \\
		\textbf{(8)}
		Return the rate estimates $(\hat{r}^c_F,\hat{r}_D,\hat{r}_R,\hat{r}_B)$.
	}
\end{algorithm}
\noindent eliminating the influence of $r_F$ entirely. This suggests a two step approach where $r_F$ is treated separately from $(r_D,r_R,r_B)$.

\subsection{Estimating parameters}
The estimation procedures consist roughly of two phases: estimation of $\eta$, and  estimation of the kinetic rates. The idea is that once $\hat{\eta}$ is known, we can obtain \textsl{location invariant} statistics, that allow estimation of the kinetic rates. The second phase is further divided into two steps, as $r_F$ is treated separately from the remaining rates.

Estimating $\eta$ is easy when $E$ is observed separately, since
\begin{equation}
\eta = 1-\frac{\intf{E}}{\intf{O}},
\end{equation}
so the problem reduces to intensity estimation, which is routinely performed by setting the observed number of points in relation to the area of the observation window. Next, to estimate the kinetic rates, the primary ingredients are the quantities
\begin{equation}
\zeta_u = \left(\gamma_1(f_u)-\gamma_2(f_u)\right)n_c, \quad u\in T,
\end{equation}
which can be extracted from the data using the identity in \Cref{IBCpp::eta}, which states that
\begin{equation}
\zeta_u = \frac{\left(\gamma_2^O(f)\mcor{O}{f}(r)\pcf{O}(r)-\indmark{f}\left[\pcf{O}(r)-1\right]-\gamma_2^O(f)\right)\intf{O}}{(\dispdense*\dispdense)(r)\sigfrac}, \quad u\in T,
\end{equation}
which is estimable on the basis of the observed data and $\hat{\eta}$. From the collection $\{\hat{\zeta}_u\}_{u \in T}$ we set up a weighted minimization problem
\begin{equation}
\min_{\hat{r}_D,\hat{r}_R,\hat{r}_B} \sum_{u \in T}\sum_{r \in R} \left(\frac{\hat{\zeta}_u}{\hat{\gamma}_2(f_u)}\right)^2\left(\hat{\zeta}_u-(\gamma_1(f_u)-\hat{\gamma}_2(f_u))n_c\right)^2,
\end{equation}
over the involved rates, where $R$ is a set of spatial distances that must be specified, and the weights $\frac{\hat{\zeta}_u}{\hat{\gamma}_2(f_u)}$ are chosen to put more weight on temporal distances that are most informative. The rates control the values of $\gamma_1(f_u)$ and $n_c$, and expressions for these are available in supplementary \hyperref[suppsec]{Section B}. As the minimization leads only to 3 of the 4 rates, $r_F$ is obtained separately via
\begin{equation}
\hat{r}_F = \left(\frac{\frac{1}{N}\sum_{i = 1}^{N}t_{o_i}-(1-\hat{\eta})\frac{b}{2} }{\hat{\eta}} -\hat{A}_2-\hat{B}_2\right)^{-1},
\end{equation}
where $\hat{A}_2$ and $\hat{B}_2$ are statistics computed on the basis of $(\hat{r}_D,\hat{r}_R,\hat{r}_B)$. Since we only observe a finite recording of lenght $b$, $\hat{r}_F$ will be subject to a censoring bias. A corrected estimate is found by solving
\begin{equation}
\frac{e^{r^c_F b}-r^c_F b-1}{r^c_F(e^{r^c_Fb}-1)}-\hat{r}_F^{-1} = 0,
\end{equation}
in $r^c_F$.

\begin{figure}
	\centering
	\includegraphics[width=1\linewidth]{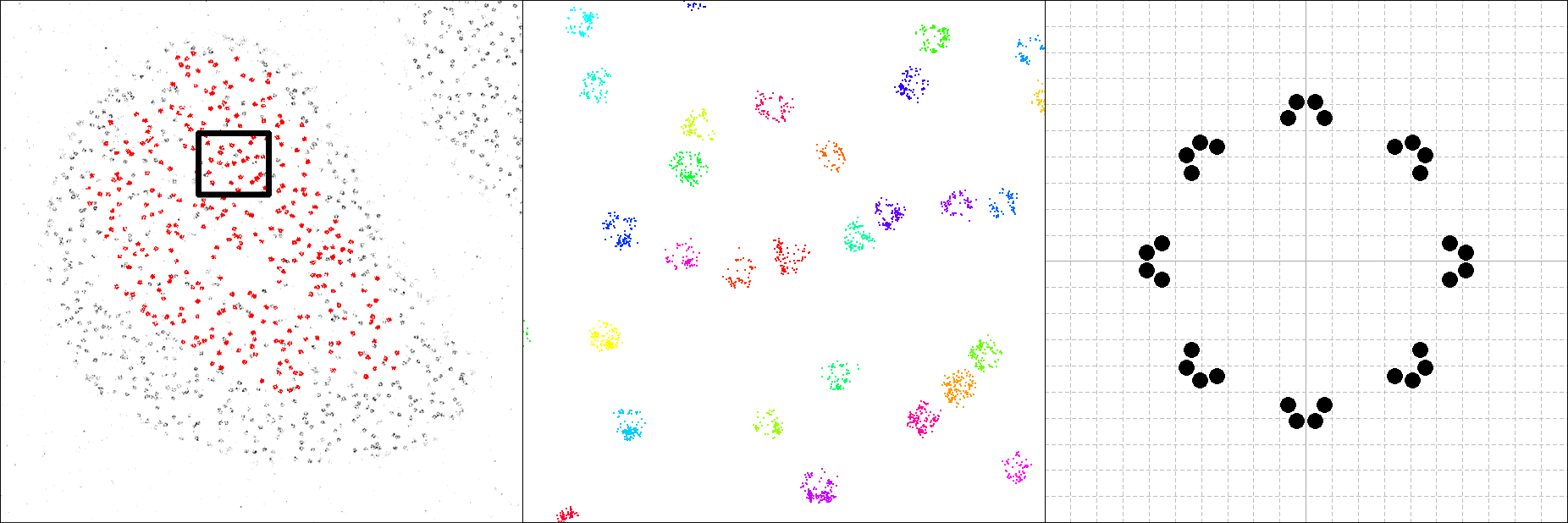}
	\caption{Nuclear pore complexes in Nup96 cell lines. Left: an example dataset of a cell expressing Nup96-mMaple. The red complexes are those that were confidently segmented by SMAP (see main text), and were used for further analysis. The rectangular region is picked only for visualization purposes, and can be seen magnified in the center plot. Center: magnified region of segmented complexes. The color indicates which points are determined as belonging to the same complex. Right: Top view schematic of an idealized Nup96 complex. A grid of separation $10$nm is overlaid for scale.}
	\label{fig:NUP}
\end{figure}

\section{Validation of methods on a nuclear pore complex reference cell line} \label{sec:NUP}
The nuclear pore complex (NPC) is quickly becoming a reference standard for quantitative SMLM imaging. In a recently developed NPC cell line \citep{Thevathasan2019}, the nucleoporin Nup96 is endogenously tagged with fluorescent labels. Each complex forms a ring of approximately $55$nm in radius, comprising $32$ Nup96 arranged into $8$ equally spaced corners of $4$ Nup96 each, see \Cref{fig:NUP}. Due to this well-characterized organization of proteins, these cells offer the rare opportunity of checking results against a known ground truth on a real biological sample. 

We analyze the publicly available datasets \citep{JervisVermalThevathasan2019} comprising PALM recordings of Nup96 tagged with mMaple, using a buffer of 50mM Tris in D$_2$O, recorded with a camera integration length of $\Delta = \frac{1}{10}$. In total, this amounts to localized data from $16$ cells, preprocessed according to the procedures in \citep{Thevathasan2019}; briefly, using the provided open-source software SMAP \citep{Ries2020}, localizations were corrected for drift, and emitters with large uncertainty or poor fit likelihood were filtered out. This data presents a challenging setting for the PALM-IBCpp analysis, as the filtering steps are a clear breach of model assumptions, and the low framerate of $10$hz challenges the approximations used in fitting, which are only exact in the limit of large framerates.

For each cell, we used SMAP with the established procedures to first segment out high-quality NPCs, and then estimated the effective labeling efficacy (ELE), which describes the fraction of Nup96 that are sufficiently bright to be detected in the SMLM recording. For each cell, we then computed the ''target'' number of reappearances per Nup96 ($\mathds{E}_S[G]$, ''$S$'' indicating SMAP) according to the formula
\begin{align}
\mathds{E}_S[G] &= \frac{N_{\mathrm{loc}}}{N_{{\mathrm{NPC}}}\cdot \mathrm{ELE}\cdot 32},
\end{align}
where $N_{{\mathrm{NPC}}}$ is the number of segmented NPC, and $N_{\mathrm{loc}}$ is the total number of localizations observed across all segmented complexes. In addition, the target number of $F$ state visits ($\mathds{E}_S[N_b]$) is computed as
\begin{align}
\mathds{E}_S[N_b] &= \frac{N_{\mathrm{loc,grouped}}}{N_{{\mathrm{NPC}}}\cdot \mathrm{ELE}\cdot 32},
\end{align}
where $N_{\mathrm{loc,grouped}}$ is the number of localizations from the segmented complexes, after grouping together localizations close in space (35nm) and time (1 frame), again according to the procedures of \citep{Thevathasan2019}. As the number of $F$ state visits has a Geometric distribution (starting at 1), we have $N_b \sim \mathrm{Geom}_1(p)$, where $p = \frac{r_B}{r_B+r_D}$ is the bleaching probability. An SMAP estimate of $p$ is thus naturally found via
\begin{equation}
p_S = \frac{1}{\mathds{E}_S[N_b]}.
\end{equation}
Finally, we fit the IBCpp model on the segmented NPCs. To get the most fair comparison with the SMAP targets, we set $\hat{\eta} = 1$ when fitting. This is because SMAP does not account for background localizations, and thus assumes all observations are generated by PA-FP. After fitting, we computed the estimated values of the above targets. We also include the derived statistic
\begin{equation}
N_{\mathrm{copy}} = \frac{N_{\mathrm{loc}}}{N_{{\mathrm{NPC}}}\cdot{\rmean{G}}\cdot \mathrm{ELE} },
\end{equation}
where ${\rmean{G}}$ is the estimated mean of $G$, on the basis of the PALM-IBCpp fit. $N_{\mathrm{copy}}$ has a ground truth target value of $32$, the copy number of Nup96 per complex.

While the true blinking rates of the data remain unknown, and have no direct SMAP analogue, we can nevertheless compare our model predictions on the derived blinking statistics against the targets, and in this way validate important aspects of our modeling and estimation framework. In \Cref{table:nupmaintab} the means and standard deviations from fitting to the 16 datasets can be seen. Interestingly, in spite of the model violations incurred by data filtering, we obtained encouraging results. The most intuitive reference quantity, $N_{\mathrm{copy}}$, is estimated at $32.3 \pm 1.82$, in close correspondence with the ground truth value of $32$. The accurate recovery is due to the tight control on $\rmean{G}$, the total number of appearances per Nup96, estimated at $7.40 \pm 0.72$ by our model, versus $7.46$ for the SMAP analysis. One slight deviation from the targets is the number of $F$ state visits, estimated at $2.32 \pm 0.08$ versus $2.93$ for SMAP. A possible explanation for this difference lies in how SMAP estimates it; since the grouping procedure only looks for repeat localizations within a spatial radius of $35$nm, it should be expected that some $F$ state visits are broken up into multiple subsegments, potentially biasing results in favor of larger values. This would also explain the slight disagreement for $p$. 

As mentioned, we unfortunately do not have an SMAP reference for the blinking rates. Nevertheless, as both the total number of reapperances and number of blinking cycles are well-estimated, it seems plausible that the estimated blinking dynamics as a whole can be trusted. Looking at the rates, we see that there is surprisingly low variability between datasets, indicating that the replications were performed with careful attention to the experimental conditions. In addition, we notice quite a long-lived dark state, lasting on average $3$ seconds. Using the mean rates across all 16 datasets, we find that the Nup96-mMaple had a mean bleaching time of $4.61$ seconds, and $99$\% of Nup96 bleached within $31$ seconds. 

Results from each individual dataset, including the ELE, number of NPC, and dataset ids, are also available, see \Cref{table:nupcomplete}. Although not used in this analysis, we also included estimates of $\eta$ for completeness.

\begin{table}[ht]\label{table:nupmain}
	\caption{Estimates and standard deviations of blinking rates and derived statistics on the basis of our model fit to 16 datasets of Nup96 NPC. The target values are based on the SMAP analysis, or are known in the case of $N_{copy}$ (see main text).}
	\label{table:nupmaintab}
	\centering
	\adjustbox{max width=\textwidth}{
		\begin{tabular}{@{\extracolsep{65pt}}lrrr}
			\toprule   
			& $Estimate$ & $Target$ & $Sd$\\ 
			\midrule		        
			$r_F \cdot 10^3$ &0.73&-&0.29 \\
			$r_B$ 		  	 &2.00&-&0.28 \\
			$r_D$ 		  	 &2.64&-&0.37 \\
			$r_R$ 		  	 &0.32&-&0.06 \\				
			\midrule			
			$N_{\mathrm{copy}}$ 	 	 &32.30&32.00&1.82 \\
			$\rmean{G}$   	 &7.40&7.46&0.72 \\
			$\rmean{N_b}$	 &2.32&2.93&0.08 \\
			$p$				 &0.43&0.34&0.01 \\					
			\bottomrule
		\end{tabular}	
	}
\end{table}
\begin{table}[ht]
	\caption{Results from 16 datasets of Nup96 nuclear pore complexes. The data ids allow identification of the exact dataset analyzed, as stored on the BioImage Archive \citep{JervisVermalThevathasan2019}. The estimated rates can be seen in columns 2 through 5. $\rmean{G}$ and $\rmean{N_b}$ are the estimated total number of reappearances per Nup96 and number of $F$ state visits on the basis of the PALM-IBCpp model fit, and $\mathds{E}_{S}[G]$ and $\mathds{E}_S[N_b]$ are the associated targets, on the basis of the SMAP analysis. ELE is the estimated fraction of Nup96 that are detectable in the dataset, as determined by SMAP. $N_{\mathrm{copy}}$ is the PALM-IBCpp estimated number of Nup96 per NUP complex, after accounting for the ELE, which has a target value of $32$. $N_{\mathrm{NPC}}$ is the number of segmented complexes. Finally, $\eta$ is the estimated fraction of non-noise points.}
	\label{table:nupcomplete}
	\centering
	\adjustbox{max width=\textwidth}{
		\begin{tabular}{@{\extracolsep{0pt}}lrrrrrrrrrrrr}
			\toprule   
			Data id&$r_{F}\cdot 10^3$ & $r_B$ & $r_D$ & $r_R$ & $\rmean{G}$ & $\mathds{E}_{S}[G]$ & $\rmean{N_b}$ & $\mathds{E}_S[N_b]$ & $N_{\mathrm{copy}}$ & $\mathrm{ELE}$ & $N_{\mathrm{NPC}}$ & $\eta$ \\
			\midrule
			181123\_6 & 0.90 & 2.28 & 2.91 & 0.30 & 6.64 & 6.43 & 2.28 & 2.64 & 31.0 & 0.45 & 313  & 0.98\\
			181123\_7 & 0.23 & 1.91 & 2.87 & 0.34 & 7.72 & 6.91 & 2.51 & 2.85 & 28.6 & 0.60 & 239  & 1.00\\
			181123\_8 & 0.67 & 2.00 & 2.51 & 0.30 & 7.25 & 6.69 & 2.26 & 2.75 & 29.5 & 0.57 & 179  & 0.99\\
			190110\_1 & 0.37 & 1.71 & 2.10 & 0.30 & 8.07 & 8.20 & 2.23 & 3.06 & 32.5 & 0.65 & 184  & 0.97\\
			190110\_2 & 0.40 & 1.81 & 2.33 & 0.53 & 7.76 & 8.20 & 2.28 & 3.05 & 33.8 & 0.60 & 420  & 0.96\\
			190111\_10& 0.75 & 1.81 & 2.54 & 0.32 & 7.89 & 7.78 & 2.40 & 3.06 & 31.6 & 0.65 & 713  & 0.97\\
			190111\_11& 0.63 & 1.69 & 2.37 & 0.33 & 8.30 & 8.15 & 2.40 & 3.16 & 31.4 & 0.63 & 846  & 0.95\\
			190111\_9 & 0.48 & 1.71 & 2.43 & 0.33 & 8.25 & 8.26 & 2.42 & 3.22 & 32.0 & 0.64 & 1080 & 0.97\\
			190118\_12& 0.59 & 2.18 & 3.08 & 0.32 & 6.97 & 6.84 & 2.41 & 2.85 & 31.4 & 0.60 & 1040 & 0.99\\
			190118\_13& 0.94 & 2.23 & 2.81 & 0.30 & 6.72 & 6.66 & 2.26 & 2.77 & 31.7 & 0.61 & 567  & 0.98\\
			190118\_14& 0.63 & 2.33 & 3.03 & 0.30 & 6.56 & 7.21 & 2.30 & 2.89 & 35.1 & 0.57 & 648  & 0.98\\
			190123\_3 & 1.37 & 2.32 & 2.96 & 0.30 & 6.56 & 7.09 & 2.27 & 2.72 & 34.6 & 0.55 & 207  & 0.98\\
			190123\_4 & 1.25 & 2.36 & 3.03 & 0.30 & 6.49 & 7.02 & 2.28 & 2.88 & 34.6 & 0.58 & 303  & 0.96\\
			190123\_5 & 0.71 & 2.37 & 3.17 & 0.31 & 6.54 & 7.14 & 2.34 & 2.86 & 34.9 & 0.60 & 578  & 0.96\\
			190502\_15& 0.86 & 1.62 & 2.02 & 0.29 & 8.39 & 8.55 & 2.24 & 3.12 & 32.6 & 0.64 & 396  & 0.99\\
			190502\_16& 0.89 & 1.67 & 2.10 & 0.30 & 8.22 & 8.21 & 2.26 & 3.04 & 31.9 & 0.62 & 440  & 0.98\\
			\bottomrule
		\end{tabular}
	}
\end{table}

\section{Blinking corrected cluster analysis of LAT-mEos3.2} \label{sec:data_sect}
Cluster analysis is perhaps the most common goal of SMLM experiments, and a great deal of effort has been put towards that end. A shared complication among all such analyses is the need to deal with artificial clustering caused by blinking artifacts, and most methods require the data to be first pre-proccessed to correct this \citep{Khater2020}. This sort of pre-processing often relies on grouping of localizations on the basis of thresholds determined heuristically or by calibration data \citep{Annibale2011a,Annibale2011}, and can have quite variable performance \citep{Lee2012}. Other methods can deal with blinking by explicitly modeling it alongside the proteins \citep{Sengupta2011}, but this limits the analyses that can be done, and requires parametric modeling of the proteins.

To overcome the challenges of quantitative cluster analysis, we suggest estimating first the blinking dynamics directly from the dataset using the PALM-IBCpp model, and subsequently correcting the desired clustering analysis for blinking biases. To exemplify this general methodology, we devise a blinking corrected test for CSR, and demonstrate it on a Jurkat T cell expressing LAT-mEos3.2 PA-FP. The dataset was recorded using PALM at a framerate of $\Delta^{-1} = 25hz$, and was then resolved and corrected for drift using ThunderSTORM \citep{Ovesny2014}.

We base our approach on the $L(r)-r$ function, a commonly used transformation of Ripley's $K$-function \citep{Ripley1976}, which has better variance properties, and is easier to interpret. The function measures spatial clustering, with values of $L(r)-r > 0$ indicating clustering, $L(r)-r = 0$ for CSR-like behavior, and $L(r)-r < 0$ indicates repulsive behavior. To test whether a given dataset follows a prescribed null model, such as CSR, one can compare the observed $L(r)-r$ function to realizations from the null model, as obtained via simulations. This approach can be made rigorous using the class of global envelope tests \citep{Myllymaeki2016}, which produce an envelope that is global in the sense that, if the observed statistic breaches the envelope at any point, it corresponds to a significant test. 

At a first glance, we cannot apply this idea directly to our data, as the null model we are testing is not just CSR, but rather CSR observed under blinking and background noise. This means that we do in fact expect to observe large values of $L(r)-r$, even for CSR proteins, and the question is rather how large this function must be to indicate significant protein clustering. Fortunately, as we are able to estimate the blinking rates, we can perform simulations from a model that approximates the null, and get a better handle on the true clustering behavior of the proteins. Of course, as this method is based on parametric bootstrapping, the significance level of the test is only guaranteed to be at the specified level if the rates are estimated perfectly, and some care is advised when interpreting results. To ensure the level of the test is approximately as specified, we suggest using simulation - we demonstrate this below.

\begin{figure}
	\centering
	\includegraphics[width=1\linewidth]{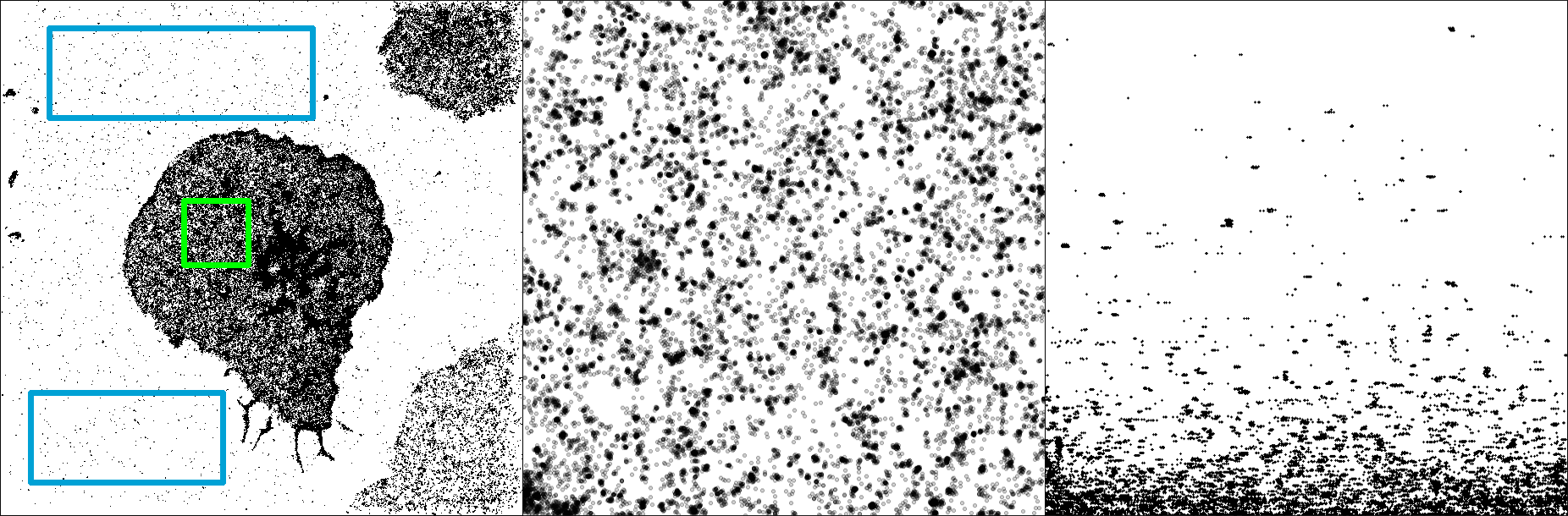}
	\caption{Left: the full dataset with green region of interest used for fitting the IBCpp model, and blue noise regions used for estimating $\eta$. Center: magnified $xy$ scatter plot of the ROI. Right: timepoints are plotted against the $x$-axis for the region of interest, demonstrating the space-time blinking dynamics.}
	\label{fig:dataset}
\end{figure}
\begin{figure}
	\centering
	\includegraphics[width=1\linewidth]{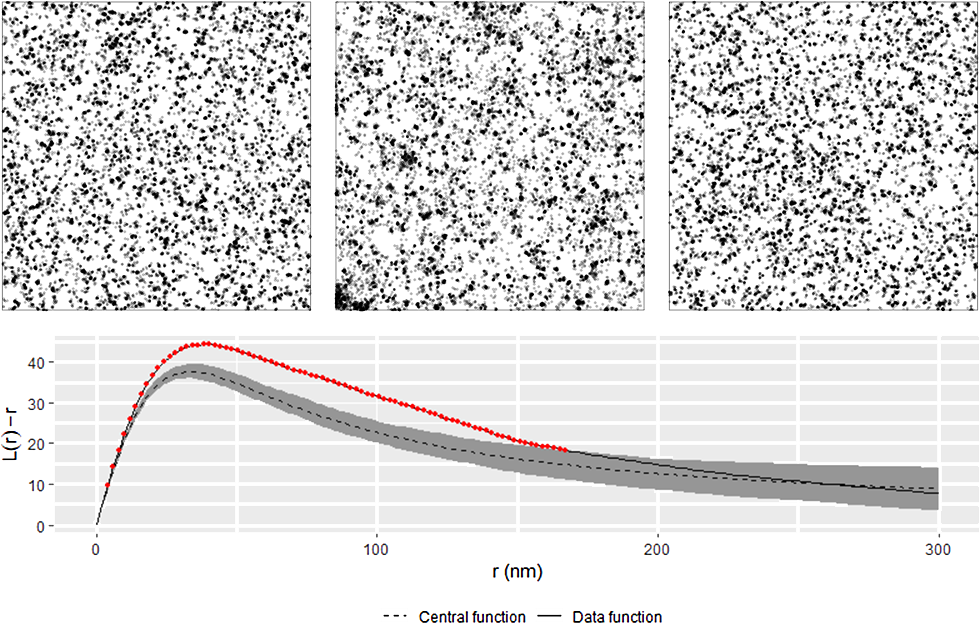}
	\caption{Blinking corrected CSR test. \textbf{Top row}. Center: the observed data. Left and right: representative simulations of blinking CSR proteins with background noise. The blinking rates, number of proteins, and noise parameters used in simulations were obtained from the PALM-IBCpp fit to the observed data (see main text). \textbf{Bottom row}. Blinking corrected, 2-sided CSR global envelope test for the observed data, on the basis of the $L(r)-r$ function. The observed $L(r)-r$ function (solid line) was compared to the $L(r)-r$ functions of $500$ simulations of blinking CSR proteins with noise, and a global envelope was constructed (shaded gray). The breach of the observed curve above the envelope indicates significant protein clustering in the ROI which cannot be explained by blinking alone ($p = 0.004$).}
	\label{fig:latcsrtest1}
\end{figure}
\begin{figure}
	\centering
	\includegraphics[width=1\linewidth]{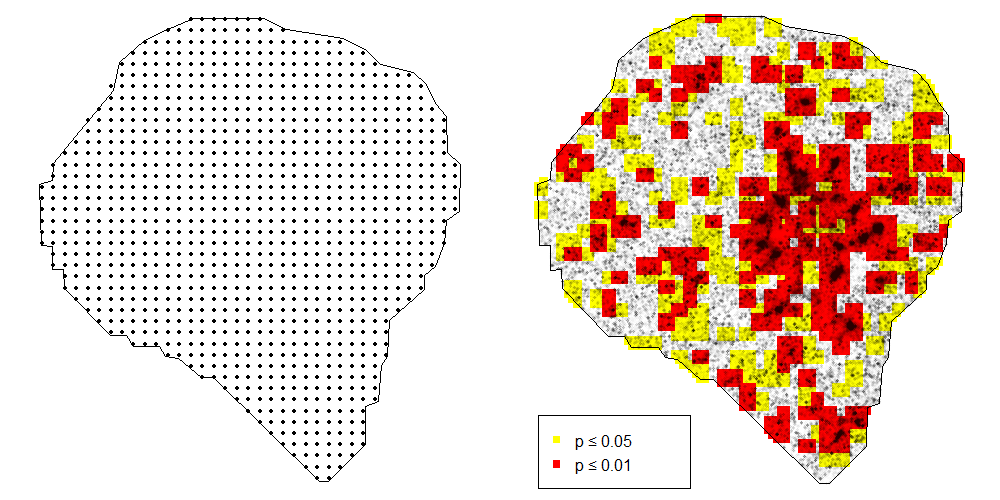}
	\caption{Blinking corrected CSR testing on the entire cell. Left: the interior of the cell was segmented out, and an evenly spaced grid with separation $500$nm was overlaid. Right: around each gridpoint, a centered $1000\times 1000$nm observation window was used to subset out a local portion of the data, and a blinking corrected CSR test was performed for that region (see main text and \Cref{fig:latcsrtest1}). For visualization purposes we extrapolated p-values to the entire cell using the p-value associated with the nearest-neighbor point in the grid.}
	\label{fig:latcsrtest2}
\end{figure}



For the analysis we first subset out a region of interest (ROI) of manageable size. In addition to the ROI we also subset out 2 large regions from the coverslip outside the cell, which were used for estimation of $\eta$, see \Cref{fig:dataset}. The ROI had $21742$ points $\{(o_k,t_{o_k})\}_{k = 1}^{21742}$ with associated localization uncertainties $\{\hat{\sigma_k}\}_{k = 1}^{21742}$. Similarly, the noise regions had $1063$ points in total, and the fraction of non-noise points (per area) was estimated at $\hat{\eta} = 0.995$. We fit the PALM-IBCpp model to this ROI, and we are thus in a position to simulate from the CSR (with blinking) null model, using the estimated blinking dynamics. To do this, the number of proteins to simulate was first determined on the basis of \Cref{mom2}, which states that
\begin{equation}
\intf{X} = \frac{\intf{Z}}{\rmean{G}} = \frac{\eta \intf{O}}{\rmean{G}},
\end{equation}
so that by plugging in our estimates for $\eta$, $\intf{O}$, and $\rmean{G}$, and multiplying by the window area, we get the number of proteins at
\begin{equation}
N_{\mathrm{protein}} \approx \frac{0.995\cdot 21742}{8.16} \approx 2651.
\end{equation}
Each localization in the blinking clusters was then simulated by adding Gaussian noise around the position of a protein, with a standard deviation sampled from $\{\hat{\sigma_k}\}_{k = 1}^{21742}$, and the timepoints were simulated according to the discretized 4 state model. Finally we added $109$ Poisson background noise points, as indicated by $\hat{\eta}$. Examples of simulations can be seen in the top row of \Cref{fig:latcsrtest1} on the left and right. 

Using this simulation scheme, we tested for CSR proteins on the basis of the $L(r)-r$ function. We computed $L(r)-r$ for the observed ROI, and obtained $500$ realizations of it from the CSR null model via simulation. We then performed a global envelope test, see \Cref{fig:latcsrtest1}. The envelopes indicate the sort of clustering that we would expect to see from blinking clusters. The observed $L(r)-r$ breaches above the envelope, indicating that there is significant clustering of proteins ($p = 0.004$). The observed ROI has spots of clustering that, upon visual inspection and comparison with the null model simulations, are clearly too large to be blinking alone. The results of fitting to the ROI can be seen in \Cref{table:datafit}, where also the results of refitting to $100$ simulations of the CSR null model are included. The refits indicate approximate unbiasedness, and low uncertainty of rate estimates. To validate that our test is approximately at the $5\%$ level,  we performed the CSR test for each of the $100$ simulations, resulting in $3$ rejections, in close correspondence with expectations.

To complete this analysis, we next performed the CSR test on the entire cell by means of a rolling window, see \Cref{fig:latcsrtest2}. This revealed regions of strongly significant clustering, but also regions indistinguishable from CSR. In fact roughly half the cell presented as clustered, with $47\%$ of the cell clustered at the $5\%$ percent level, and $26\%$ at the $1\%$ level. 

\begin{table}[ht]
	\caption{Estimates (Est) obtained from the fit to the observed data. Included is average (Avg) and standard deviation (Sd) of estimates obtained from fitting to $100$ simulations from the CSR null model. Included derived quantities are: the mean number of appearances per protein, $\rmean{G}$, the bleaching probability, $p = \frac{r_B}{r_B+r_D}$, and the $(0.25,0.50,0.75,0.99)$-quantiles $(q_{0.25},q_{0.50},q_{0.75},q_{0.99})$ of the total PA-FP lifetime distribution (time in seconds from activation to bleaching). For example, $75\%$ of PA-FP bleach within $q_{0.75}$ seconds.}
	\label{table:datafit}
	\centering
	\adjustbox{max width=\textwidth}{
		\begin{tabular}{@{\extracolsep{65pt}}lrrrr}
			\toprule   
			& Est & Avg & Sd\\ 
			\midrule		   		
			$r_{F}\cdot 10^{3}$ & 5.16 &  5.17 & 0.13 \\ 
			$r_{B}$& 4.92 &  5.07 & 0.13 \\
			$r_{D}$& 10.50 & 11.40 & 0.55 \\
			$r_{R}$& 1.11 &  1.15 & 0.04 \\				
			\midrule
			$\rmean{G}$& 8.16 &  8.13 & 0.17 \\
			$p$& 0.32 &  0.31 & 0.01 \\
			$q_{0.25}$& 0.10 &  0.10 & 0.00 \\
			$q_{0.50}$& 1.04 &  1.08 & 0.04 \\
			$q_{0.75}$& 3.10 &  3.13 & 0.05 \\
			$q_{0.99}$& 15.70 & 12.70 & 0.15 \\
			\bottomrule
		\end{tabular}	
	}
\end{table}

\begin{figure}
	\centering
	\includegraphics[width=1\linewidth]{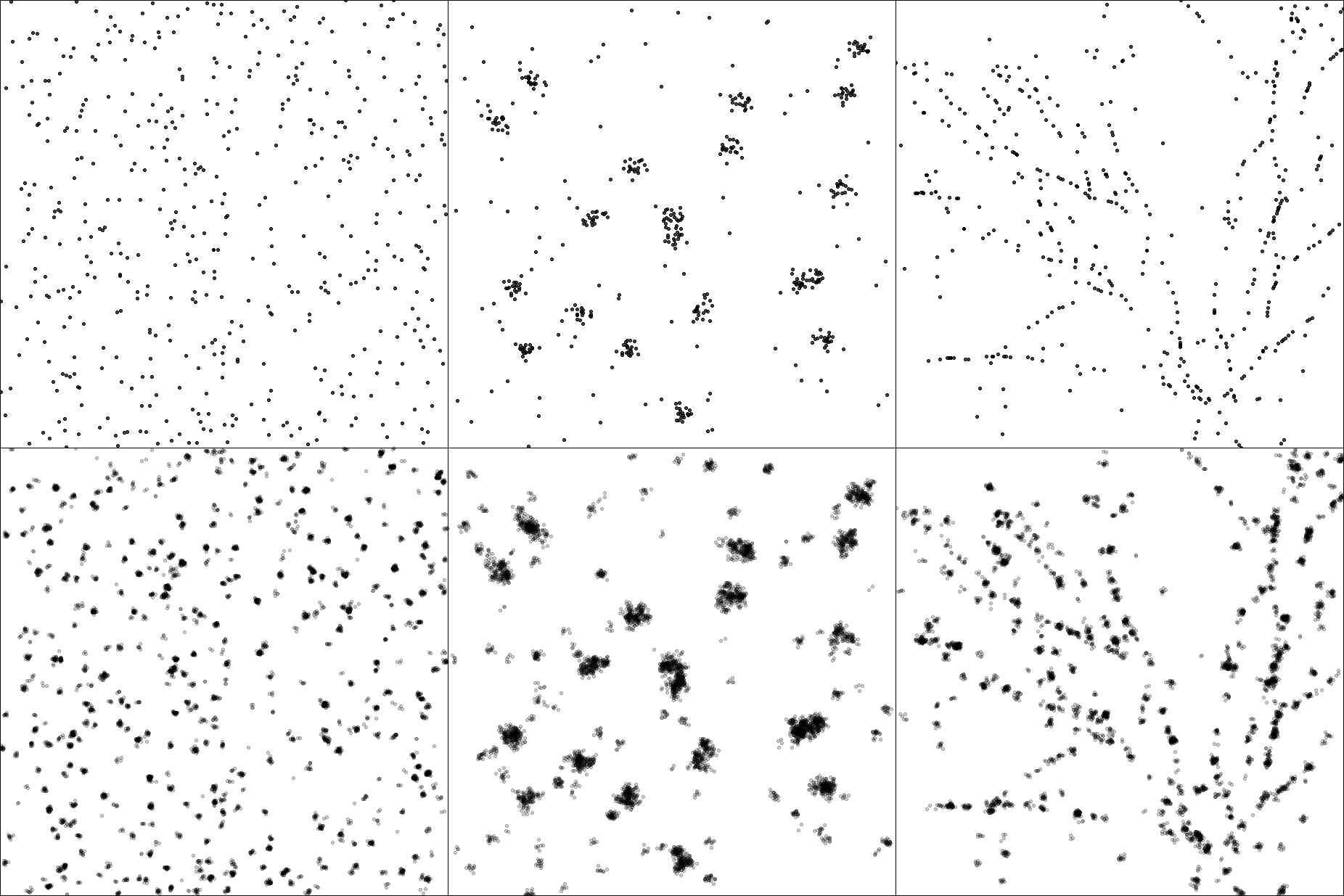}
	\caption{Typical simulations from each of the 3 protein configurations (CSR, clusters, fibers) in the columns, before and after adding blinking clusters in the top and bottom rows, respectively. The CSR data is simulated as 500 i.i.d. uniform points in the ROI. The clusters data consists of 100 CSR points and further 20 uniformly located Gaussian clusters with standard deviation $50$, each having 20 points. Finally, for the fiber data, 450 points are sampled uniformly along the edges of a fixed fiber structure, and 50 CSR points are added to the background.}
	\label{fig:simex}
\end{figure}
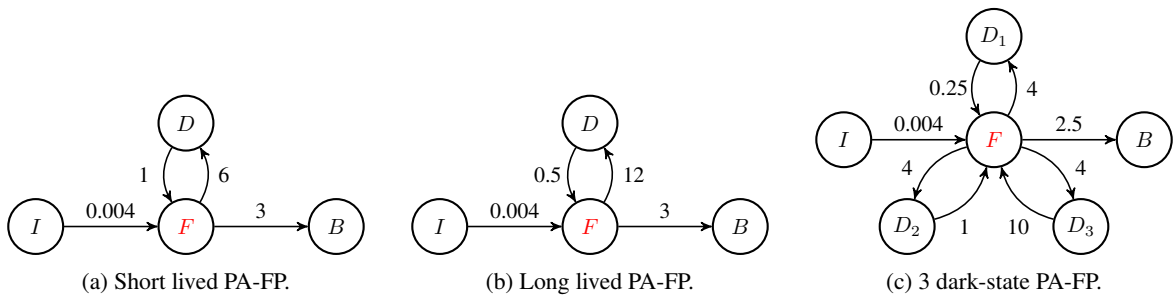
\begin{figure}
	\centering
	\begin{subfigure}[t]{0.32\textwidth}
		\centering
		\begin{tikzpicture}[->, >=stealth', auto, semithick, node distance=1.5cm, 
		scale=0.83, every node/.append style={transform shape}]
		\tikzstyle{every state}=[fill=white,draw=black,thick,text=black,scale=1]
		\node[state]    (F)  {\textcolor{red}{$F$}};
		\node[state]    (I)[left = of F]  {$I$};
		\node[state]    (B)[right = of F]  {$B$};
		\node[state]    [above = 0.75cm of F] (D)  {$D$};
		\path
		(I) edge[]     node{0.004}     (F)
		(F) edge[bend right]     node{\hspace{-2cm}1}     (D)
		(D) edge[bend right]     node{\hspace{0.75cm}6}     (F)
		(F) edge[]     node{3}     (B);
		\end{tikzpicture}
		\caption{Short lived PA-FP.}
	\end{subfigure}
	\begin{subfigure}[t]{0.32\textwidth}
		\centering
		\begin{tikzpicture}[->, >=stealth', auto, semithick, node distance=1.5cm, 
		scale=0.83, every node/.append style={transform shape}]
		\tikzstyle{every state}=[fill=white,draw=black,thick,text=black,scale=1]
		\node[state]    (F)  {\textcolor{red}{$F$}};
		\node[state]    (I)[left = of F]  {$I$};
		\node[state]    (B)[right = of F]  {$B$};
		\node[state]    [above = 0.75cm of F] (D)  {$D$};
		\path
		(I) edge[]     node{0.004}     (F)
		(F) edge[bend right]     node{\hspace{-2cm}0.5}     (D)
		(D) edge[bend right]     node{\hspace{0.75cm}12}     (F)
		(F) edge[]     node{3}     (B);
		\end{tikzpicture}
		\caption{Long lived PA-FP.}
	\end{subfigure}
	\begin{subfigure}[t]{0.32\textwidth}
		\centering
		\begin{tikzpicture}[->, >=stealth', auto, semithick, node distance=1.5cm, 
		scale=0.83, every node/.append style={transform shape}]
		\tikzstyle{every state}=[fill=white,draw=black,thick,text=black,scale=1]
		\node[state]    (F)  {\textcolor{red}{$F$}};
		\node[state]    (I)[left = of F]  {$I$};
		\node[state]    (B)[right = of F]  {$B$};
		\node[state]    [above = 0.75cm of F] (D1)  {$D_1$};
		\node[state]    [below left  = 0.75cm and 0.75cm of F] (D2)  {$D_2$};
		\node[state]    [below right = 0.75cm and 0.75cm of F] (D3)  {$D_3$};
		\node[]    [above = 0.25cm of D3]  {4};
		\node[]    [above = 0.25cm of D2]  {4};
		\node[]    [right = 0.25cm of D2]  {1};
		\node[]    [left = 0.25cm of D3]  {10};
		\path
		(I) edge[]     node{0.004}     (F)
		(F) edge[bend right]     node{\hspace{-2.1cm}0.25}     (D1)
		(D1) edge[bend right]     node{\hspace{0.75cm}4}     (F)
		(F) edge[bend right]     node{}     (D2)
		(F) edge[bend left]     node{}     (D3)
		(D2) edge[bend right]     node{}     (F)
		(D3) edge[bend left]     node{}     (F)
		(F) edge[]     node{2.5}     (B);
		\end{tikzpicture}
		\caption{3 dark-state PA-FP.}
	\end{subfigure}
	\caption{The three models of PA-FP photophysics considered in simulations.}
	\label{fig:PAFPM}
\end{figure}
\begin{table}
	\caption{Results of fitting to 100 simulations from each combination of protein distribution (CSR, cluster, fibers) and PA-FP model (short lived, long lived, 3 dark-states). The average (Avg) and standard deviation (Sd) of estimates is included.}
	\label{table:Simfits}
	\centering
	\adjustbox{max width=\textwidth}{
		\begin{tabular}{@{\extracolsep{5.5pt}}lrrrrrrrrrrrr}
			\multicolumn{12}{c}{CSR} \\
			\toprule   
			{} & \multicolumn{3}{c}{Short lived} & {} & \multicolumn{3}{c}{Long lived} &{} & \multicolumn{3}{c}{3 dark-states}\\
			\cmidrule{2-4} 
			\cmidrule{6-8} 
			\cmidrule{10-12} 
			& Truth & Avg & Sd & & Truth & Avg & Sd & & Truth & Avg & Sd\\ 
			\midrule
			$r_{F}\cdot 10^3$ & 4.00 & 3.98 & 0.23 && 4.00  & 4.04  & 0.26 && 4.00 & 3.98 & 0.23 \\
			$r_{B}$ 		  & 3.00 & 3.10 & 0.20 && 3.00  & 3.15  & 0.27 && 2.50 & 2.31 & 0.19 \\
			$r_{D}$		  	  & 6.00 & 6.59 & 0.67 && 12.00 & 13.40 & 1.22 &&      & 7.08 & 0.56 \\
			$r_{R}$			  & 1.00 & 1.08 & 0.09 && 0.50  & 0.54  & 0.05 &&      & 0.44 & 0.06 \\
			\midrule
			$\rmean{G}$		  & 11.30 & 11.20 & 0.68  && 13.25 & 13.20  & 0.83 && 15.38 & 14.97 & 1.04 \\
			$p$ 			  & 0.33  & 0.32  & 0.02  && 0.20  & 0.19   & 0.01 && 0.17  & 0.25  & 0.01 \\
			$q_{0.25}$ 		  & 0.16  & 0.16  & 0.02  && 0.76  & 0.89   & 0.18 && 0.44  & 0.39  & 0.11 \\
			$q_{0.50}$ 		  & 1.12  & 1.17  & 0.16  && 4.92  & 5.01   & 0.56 && 4.16  & 4.27  & 0.69 \\
			$q_{0.75}$ 		  & 3.36  & 3.35  & 0.33  && 12.00 & 12.00  & 1.26 && 12.20 & 11.10 & 1.68 \\
			$q_{0.99}$ 		  & 13.80 & 13.50 & 1.17  && 45.50 & 44.70  & 4.53 && 50.00 & 42.82 & 6.32 \\
			\bottomrule
		\end{tabular}  
	}
	\adjustbox{max width=\textwidth}{ 
		\begin{tabular}{@{\extracolsep{5.5pt}}lrrrrrrrrrrrr}
			\multicolumn{12}{c}{Clusters} \\
			\toprule   
			{} & \multicolumn{3}{c}{Short lived} & {} & \multicolumn{3}{c}{Long lived} &{} & \multicolumn{3}{c}{3 dark-states}\\
			\cmidrule{2-4} 
			\cmidrule{6-8} 
			\cmidrule{10-12} 
			& Truth & Avg & Sd & & Truth & Avg & Sd & & Truth & Avg & Sd\\ 
			\midrule
			$r_{F}\cdot 10^3$ & 4.00 & 4.00 & 0.27 && 4.00  & 3.98  & 0.24 && 4.00 & 3.99 & 0.26 \\
			$r_{B}$ 		  & 3.00 & 3.06 & 0.20 && 3.00  & 3.10  & 0.26 && 2.50 & 2.29 & 0.23 \\
			$r_{D}$ 		  & 6.00 & 6.53 & 0.62 && 12.00 & 13.10 & 1.39 &&      & 6.97 & 0.63 \\
			$r_{R}$ 		  & 1.00 & 1.06 & 0.09 && 0.50  & 0.53  & 0.06 &&      & 0.43 & 0.06 \\
			\midrule
			$\rmean{G}$ 	  & 11.30 & 11.30 & 0.69 && 13.25 & 13.30 & 0.79 && 15.38 & 15.03 & 1.23 \\
			$p$ 			  & 0.33  & 0.32  & 0.02 && 0.20  & 0.19  & 0.01 && 0.17  & 0.25  & 0.02 \\
			$q_{0.25}$ 		  & 0.16  & 0.16  & 0.02 && 0.76  & 0.90  & 0.16 && 0.44  & 0.38  & 0.11 \\
			$q_{0.50}$ 		  & 1.12  & 1.19  & 0.18 && 4.92  & 5.09  & 0.52 && 4.16  & 4.33  & 0.74 \\
			$q_{0.75}$ 		  & 3.36  & 3.41  & 0.39 && 12.00 & 12.30 & 1.23 && 12.20 & 11.30 & 1.80 \\
			$q_{0.99}$ 		  & 13.80 & 13.70 & 1.37 && 45.50 & 45.50 & 4.56 && 50.00 & 43.68 & 6.82 \\
			\bottomrule
		\end{tabular} 
	}
	\adjustbox{max width=\textwidth}{
		\begin{tabular}{@{\extracolsep{5.5pt}}lrrrrrrrrrrrr}
			\multicolumn{12}{c}{Fibers} \\
			\toprule   
			{} & \multicolumn{3}{c}{Short lived} & {} & \multicolumn{3}{c}{Long lived} &{} & \multicolumn{3}{c}{3 dark-states}\\
			\cmidrule{2-4} 
			\cmidrule{6-8} 
			\cmidrule{10-12} 
			& Truth & Avg & Sd & & Truth & Avg & Sd & & Truth & Avg & Sd\\ 
			\midrule
			$r_{F}\cdot 10^3$ & 4.00 & 4.01 & 0.25 && 4.00  & 4.03  & 0.24 && 4.00 & 4.02 & 0.24 \\
			$r_{B}$ 		  & 3.00 & 3.11 & 0.21 && 3.00  & 3.21  & 0.30 && 2.50 & 2.28 & 0.20 \\
			$r_{D}$ 		  & 6.00 & 6.66 & 0.64 && 12.00 & 13.60 & 1.29 &&      & 7.02 & 0.58 \\
			$r_{R}$ 		  & 1.00 & 1.09 & 0.12 && 0.50  & 0.54  & 0.05 &&      & 0.43 & 0.06 \\
			\midrule
			$\rmean{G}$ 	  & 11.30 & 11.20 & 0.66 && 13.25 & 13.10 & 0.93 && 15.38 & 15.13 & 1.12 \\
			$p$ 			  & 0.33 & 0.32   & 0.02 && 0.20 & 0.19   & 0.01 && 0.17  & 0.25  & 0.01 \\
			$q_{0.25}$ 		  & 0.16 & 0.16   & 0.02 && 0.76 & 0.89   & 0.21 && 0.44  & 0.40  & 0.11 \\
			$q_{0.50}$ 		  & 1.12 & 1.17   & 0.15 && 4.92 & 5.01   & 0.64 && 4.16  & 4.33  & 0.69 \\
			$q_{0.75}$ 		  & 3.36 & 3.35   & 0.32 && 12.00 & 12.00 & 1.41 && 12.20 & 11.24 & 1.68 \\
			$q_{0.99}$ 		  & 13.80 & 13.50 & 1.19 && 45.50 & 44.70 & 5.01 && 50.00 & 43.34 & 6.30 \\
			\bottomrule
		\end{tabular}
	}
\end{table}

\section{Simulation study} \label{sec:simstud}
We evaluate the performance of our method under different protein distributions and blinking models. We will also consider what happens when the blinking model is misspecified. We consider 3 different cases of protein distributions: CSR, spherical clustering, and fibrous structures, see \Cref{fig:simex}. We fix the number of proteins at $500$ for all simulations, with localization uncertainties drawn i.i.d. from the Gamma(6.5, 0.375) distribution (shape and rate parameterization), which is the maximum likelihood fit to the observed uncertainties in the LAT data of \Cref{sec:data_sect}, and we consider $\eta = 1$ known. 

For the kinetic rates, we consider short and long lived PA-FPs. Additionally, in a misspecified case, we use a model with 3 distinct dark states, each selected with the same probability, but with very different holding time distributions. For the values of the kinetic rates in the 3 PA-FP models, see \Cref{fig:PAFPM}. We simulated $100$ realizations from each combination of spatial organization and blinking behavior, and discretized signals according to a framerate of 25hz.  

The results of the simulation study can be seen in \Cref{table:Simfits}. For the short and long lived PA-FPs, we see that there is close correspondence between the true parameter values and their estimates, especially for the smaller rates and all derived blinking statistics. The mean number of reappearances is well estimated, as is the bleaching probability $p$, and the total lifetime quantiles. Some bias appears to exist for the dark-state entrance rate, $r_D$, which also has the highest uncertainty of the rates. This is likely due to bias in the utilized approximations for low framerate to rate ratios. Importantly, for the misspecified 3 dark-states model, the number of reappearances and the lifetime quantiles are again well estimated. Unsurprisingly, both $r_B$ and $p$ are biased in this case, as the model attempts to fit to an average blinking cycle, and cannot exactly capture the nuances of having 3 different dark states. Overall, the effect of the protein distribution is small compared to the effect of different PA-FP models, with a slight increase in variance for more clustered conditions.

To put this analysis into a broader perspective, we compared with results obtained from the \textsc{PC-PALM} (pair correlation PALM) method of \citep{Sengupta2011}. This method is not capable of extracting the kinetic rates, but it can estimate $\rmean{G}$ for sufficiently simple models on the distribution of $G$. The PC-PALM method requires modeling of the proteins via an assumed form for the protein pair correlation function $\pcf{X}(r)$. Following the authors, we set
\begin{equation}
\pcf{X}(r) = Ae^{-\frac{r}{B}}+1,
\end{equation}
where $A$ and $B$ are parameters that need to be estimated. The PC-PALM method fits a model to the observed pair correlation function, from which $n_c$ is readily estimated. In order to then estimate $\rmean{G}$, the authors use the approximation
\begin{equation}
\rmean{G} \approx n_c,
\end{equation}
which, as noted by \citep{Veatch2012} and \citep{Andersen2018}, holds exactly if $G$ has a Poisson distribution. Given the 4-state model of PA-FP photophysics, we argue a Geometric distribution is more appropriate, in which case we would have
\begin{equation}
\rmean{G} = \frac{n_c}{2}+1.
\end{equation}
Using both these estimators, referred to as PC-PALM 1 and PC-PALM 2, respectively, we compared performance with the IBCpp fit on the simulated data, the results of which can be seen in \Cref{table:pcpalm}. We see that the IBCpp fit has lower bias and variance in every case, and is less sensitive to the blinking and clustering properties of PA-FP. PC-PALM is sensitive to the assumed distribution for $G$ and $\pcf{X}$, which is particularly clear in the 3 dark-state model, which has the most complex blinking behavior, and for the fibers, which has the most heterogeneous spatial distribution.

\begin{table}
	\caption{Comparison with the PC-PALM method for estimating $\rmean{G}$. Average (Avg), bias (Bias) given as the true value minus Avg, and standard deviation (Sd) of estimates is included. The first PC-PALM method assumes a Poisson distribution for $G$, whereas the second assumes a Geometric distribution. Best values are in bold.}
	\label{table:pcpalm}
	\centering
	\adjustbox{max width=\textwidth}{
		\begin{tabular}{@{\extracolsep{5.5pt}}lrrrrrrrrrrr}
			\multicolumn{12}{c}{CSR} \\
			\toprule   
			{} & \multicolumn{3}{c}{Short lived} & {} & \multicolumn{3}{c}{Long lived} &{} & \multicolumn{3}{c}{3 dark-states}\\
			\cmidrule{2-4} 
			\cmidrule{6-8} 
			\cmidrule{10-12} 
			& Avg & Bias & Sd && Avg & Bias & Sd && Avg & Bias & Sd \\
			\midrule						
			PC-PALM 1 & 21.10 & 9.80 & 1.52 &  & 25.40 & 12.15 & 1.92 &  & 41.20 & 25.82 & 4.40 \\
			PC-PALM 2 & 11.50 & 0.20 & 0.76 &  & 13.70 & 0.45 & 0.96 &  & 21.60 & 6.22 & 2.20 \\
			IBCpp & 11.20 & \textbf{-0.10} & \textbf{0.68} &  & 13.20 & \textbf{-0.05} & \textbf{0.83} &  & 15.00 & \textbf{-0.38} & \textbf{1.04} \\
			\bottomrule
		\end{tabular}  
	}
	\adjustbox{max width=\textwidth}{
		\begin{tabular}{@{\extracolsep{5.5pt}}lrrrrrrrrrrr}
			\multicolumn{12}{c}{Clusters} \\
			\toprule   
			{} & \multicolumn{3}{c}{Short lived} & {} & \multicolumn{3}{c}{Long lived} &{} & \multicolumn{3}{c}{3 dark-states}\\
			\cmidrule{2-4} 
			\cmidrule{6-8} 
			\cmidrule{10-12} 
			& Avg & Bias & Sd && Avg & Bias & Sd && Avg & Bias & Sd \\
			\midrule						
			PC-PALM 1 & 18.10 & 6.80 & 2.70 &  & 21.70 & 8.45 & 2.76 &  & 25.20 & 9.82 & 3.40 \\
			PC-PALM 2 & 10.00 & -1.30 & 1.35 &  & 11.90 & -1.35 & 1.38 &  & 13.60 & -1.78 & 1.70 \\
			IBCpp & 11.30 & \textbf{0.00} & \textbf{0.69} &  & 13.30 & \textbf{0.05} & \textbf{0.79} &  & 15.00 & \textbf{-0.38} & \textbf{1.23} \\
			\bottomrule
		\end{tabular}  
	}
	\adjustbox{max width=\textwidth}{
		\begin{tabular}{@{\extracolsep{5.5pt}}lrrrrrrrrrrr}
			\multicolumn{12}{c}{Fibers} \\
			\toprule   
			{} & \multicolumn{3}{c}{Short lived} & {} & \multicolumn{3}{c}{Long lived} &{} & \multicolumn{3}{c}{3 dark-states}\\
			\cmidrule{2-4} 
			\cmidrule{6-8} 
			\cmidrule{10-12} 
			& Avg & Bias & Sd && Avg & Bias & Sd && Avg & Bias & Sd \\
			\midrule						
			PC-PALM 1 & 25.70 & 14.40 & 1.95 &  & 30.50 & 17.25 & 2.49 &  & 36.60 & 21.22 & 2.92 \\
			PC-PALM 2 & 13.80 & 2.50 & 0.98 &  & 16.20 & 2.95 & 1.25 &  & 19.30 & 3.92 & 1.46 \\
			IBCpp & 11.20 & \textbf{-0.10} & \textbf{0.66} &  & 13.10 & \textbf{-0.15} & \textbf{0.93} &  & 15.10 & \textbf{-0.28} & \textbf{1.12} \\
			\bottomrule
		\end{tabular}  
	}
\end{table}

\section{Summary and discussion}
In the present paper we have established the IBCpp family of spatio-temporal clustered point processes, which is suitable for SMLM data, and we have provided a useful result on the mark correlation function. We constructed the PALM-IBCpp, which is an IBCpp model particularly well-suited for PALM data, and we have presented an algorithm for estimation of the blinking dynamics. The special structure of the mark correlation function in the IBCpp family allows for a semiparametric, moment-based approach to estimation, which can be carried out without having to specify a model for the proteins. The methods were validated on nuclear pore complex reference data, where we could demonstrate a close correspondence between the model fit and expected blinking targets.

To demonstrate how the PALM-IBCpp can aid cluster analysis in PALM studies, we considered a real dataset expressing the adaptor protein LAT. We devised a blinking corrected global envelope test for CSR, and demonstrated it on the LAT data. In this way we could show that roughly half of the cell was subject to significant protein clustering, while the other half was not significantly different from CSR. We also performed a refitting study, again demonstrating the ability of the PALM-IBCpp model to accurately recover blinking dynamics in a realistic setting.

The ability to obtain blinking dynamics from any given ROI, without a need for calibration data or parametric modeling of protein locations, is perhaps the most important feature of our method, as it ensures that the estimated kinetic rates are relevant to the ROI being analyzed. The well-known sensitivity of PA-FP photodynamics to the experimental conditions \citep{Annibale2011a, Staudt2020} means that kinetic rates obtained via a calibration sample may not be entirely applicable in another sample, emphasizing the importance of being able to directly estimate data artifacts from a given ROI. Another key aspect of our method is how quickly it can be carried out, even on large ROIs. Fitting to the LAT ROI in \Cref{sec:data_sect}, which consisted of $21742$ localizations, took $45$ seconds to complete, on a laptop with an Intel Core i7 Processor (4x 1.80 GHz). The RAM usage was similarly modest, requiring $1.5$GB at the peak. 

The drawbacks of our method are as follows. First, although the IBCpp family is generally applicable to SMLM data, the estimation algorithm developed here is specifically for the PALM-IBCpp, and estimation in other SMLM modalities would require additional work. The 4-state photoblinking model will be appropriate for some PA-FP, whereas it will be a surrogate model for other PA-FP with more complex blinking dynamics. As we have seen, the PALM-IBCpp fit is still able to capture important descriptors of blinking dynamics when the model is misspecified, but the parameters of the true blinking model will remain unknown. Finally, as the methods are built on a semiparametric model, and a complex set of estimation choices, theoretical results on the estimators are not forthcoming. The simulation studies suggest that the estimators are well behaved, but we can only guess at this in general.

\section*{Acknowledgements}
This work was supported by the Centre for Stochastic Geometry and Advanced Bioimaging, funded by a grant from the Villum Foundation. We acknowledge the use of the Nikon Imaging Facility (NIC) at King’s College London for data acquisition.


\bibliography{bib}

\clearpage

\begin{center}
	SUPPLEMENTARY MATERIAL
\end{center}
\numberwithin{equation}{section}
	\numberwithin{equation}{section}
	\renewcommand{\thesection}{A}
	\begin{center}
		SECTION A: MOMENT RESULTS FOR IBCPP MODELS
	\end{center}
	\section{Moment results for IBCpp models} \label{suppA}
	Let $O$ be an IBCpp with motion-invariant protein process $\ground X$. Deriving the results of Section 3.2 is perhaps most easily done by taking as starting point the \textit{$f$-weighted second factorial moment measure}, $\alpha_{f}^{(2)}$, given as
	\begin{equation}
	\alpha_{f}^{(2)}(A) = 
	\rmean{\sum_{(o_1,t_{o_1}),(o_2,t_{o_2}) \in O^2}^{\neq} \mathds{1}_{A}(o_1,o_2)f(t_{o_1}, t_{o_2})},
	\end{equation}
	for $A \in \R^d\times\R^d$ a Borel set. By use of a Cambell theorem we obtain
	\begin{equation}
	\alpha_{f}^{(2)}(A) =  \int_A \left[\int f(t_{o_1}, t_{o_2}) d\palm{2}{O}{(o_1,o_2)}(t_{o_1}, t_{o_2})\right] d\alpha^{(2)}_O(o_1,o_2),
	\end{equation}
	so that, comparing the above with the definition of the mark correlation function, we get the alternative characterization
	\begin{equation}
	\mcor{O}{f}(o_1,o_2) = \frac{1}{\int\int f(t_{o_1},t_{o_2}) d\palm{1}{O}{o_1}(t_{o_1})d\palm{1}{O}{o_2}(t_{o_2})} \frac{\partial \alpha_{f}^{(2)}}{\partial \alpha_O^{(2)}}(o_1, o_2),
	\end{equation}
	and we need merely compute the involved factors. We first compute the 1-point mark distributions. Let $A \subset \R^d$ and $B \subset \R_+$ be Borel sets, then we obtain
	\begin{align}
	\mint{O}(A\times B) 
	&= \rmean{\sum_{x \in \ground X} \sum_{(y,t_y) \in Y_{x}} \mathds{1}_{A}(y)\mathds{1}_B(t_y)}+\rmean{\sum_{(e,t_e) \in E}\mathds{1}_{A}(e)\mathds{1}_B(t_e)} \\
	&= \rmean{\sum_{x \in \ground X} \sum_{i = 1}^{G} \mathds{1}_{A}(x+\epsilon_i)\mathds{1}_B(t_{y_i})}+\rmean{\sum_{(e,t_e) \in E}\mathds{1}_{A}(e)\mathds{1}_B(t_e)} \\
	&= \frac{\rmean{\sum_{i = 1}^{G}\mathds{1}_B(t_{y_i})}}{\rmean{G}} \int_A \intf{Z} dz + \int_B \frac{\mathds{1}(t \in [0,b])}{b}dt\int_A \intf{E} de \\
	&= \int_A \left(\eta\frac{\rmean{\sum_{i = 1}^{G}\mathds{1}_B(t_{y_i})}}{\rmean{G}}
	+(1-\eta)\int_B\frac{\mathds{1}(t \in [0,b])}{b}dt \right)d\mint{O}(o), 
	\end{align}
	where we exploited that the $\epsilon_i$ are independent of $\ground X$, $G$, and $\{t_{y_i}\}_{i = 1}^G$, when going from the second to the third line. We also used the assumed form for the intensity function of $E$ in this step. From the above, we see that all involved mark distributions are independent of locations, with
	\begin{align}
	\palms{1}{Z}(B) &= \frac{\rmean{\sum_{i = 1}^{G}\mathds{1}_B(t_{y_i})}}{\rmean{G}}, \\
	\palms{1}{E}(B) &= \int_B\frac{\mathds{1}(t \in [0,b])}{b}dt, \\
	\palms{1}{O}(B) &= \eta\palms{1}{Z}(B)+(1-\eta)\palms{1}{E}(B),
	\end{align}
	based on which the normalization for $\mcor{O}{f}$ is found
	\begin{align}
	\gamma_2^O(f) &= \int\int f(t_{o_1},t_{o_2}) d\palms{1}{O}(t_{o_1})d\palms{1}{O}(t_{o_2}) \\ &= 
	\sigfrac^2\indmark{f} + 
	(1-\sigfrac)^2\gamma_2^E(f)+2(1-\sigfrac) \sigfrac\gamma_2^{EZ}(f),
	\end{align}
	where
	\begin{align}
	\gamma_2^E(f) &= \int \int f(t_1,t_2) d\palms{1}{E}(t_1)d\palms{1}{E}(t_2), \\
	\gamma_2^{EZ}(f) &= \int \int f(t_1,t_2) d\palms{1}{E}(t_1)d\palms{1}{Z}(t_2). 
	\end{align}
	Next, we consider the second factorial moment measure of the typical cluster, $\alpha_{Y_{x}}^{(2)}$, which will be needed below. For arbitrary blinking cluster $Y_x$ and Borel sets $A \subset \R^d\times\R^d$, $B \subset \R_+ \times \R_+$, we have
	\begin{align}
	\alpha_{Y_{x}}^{(2)}(A\times B) &= 
	\rmean{\sum_{(i,j) = 1}^{G} \mathds{1}(i \neq j) \mathds{1}_A(x+\epsilon_i, x+\epsilon_j)\mathds{1}_B(t_{y_i}, t_{y_j})} \\
	&=
	\rmean{\sum_{(i,j) = 1}^{G}\mathds{1}(i \neq j)\mathds{1}_B(t_{y_i}, t_{y_j})} \\ &\times \int_{A} \dispdense(x_1-x)\dispdense(x_2-x) d(x_1,x_2),\nonumber
	\end{align}
	obtained by averaging out the $\epsilon_i$ by conditioning on $(G, \{t_{y_i}\}_{i = 1}^G)$, from which follows (by observing what happens for $B = \R_+^2$)
	\begin{align}
	\alpha_{\ground Y_{x}}^{(2)}(A) &= \label{alpha2groundY}
	\rmean{G(G-1)}\int_{A} \dispdense(x_1-x)\dispdense(x_2-x) d(x_1,x_2), \\
	\palms{2}{Y_{x}} &= \frac{\rmean{\sum_{(i,j) = 1}^{G}\mathds{1}(i \neq j)\mathds{1}_B(t_{y_i}, t_{y_j})}}{\rmean{G(G-1)}} \label{palm2Y},
	\end{align}
	and we see that $\palms{2}{Y_{x}}$ is independent of $x$. Finally, for the density of $\alpha_{f}^{(2)}$, we split the summation according to the process memberships of each pair (respectively, two points from the same cluster, points from different clusters, one cluster and one noise point, two noise points):
	\begin{align}
	\alpha_{f}^{(2)}(A) &= 
	\rmean{\sum_{x \in \ground X} \sum_{(i,j) = 1}^{G} \mathds{1}(i \neq j)\mathds{1}_{A}(x+\epsilon_i, x_2+\epsilon_j)f(t_{y_i}, t_{y_j})} \\
	&+
	\rmean{\sum_{(x_1,x_2) \in \ground X^2}^{\neq} \sum_{i = 1}^{G}\sum_{j = 1}^{G'}\mathds{1}_{A}(x_1+\epsilon_i, x_2+\epsilon'_j)f(t_{y_i}, t'_{y_j})} \nonumber\\
	&+
	\rmean{\sum_{(z,t_z) \in Z} \sum_{(e,t_e) \in E}\mathds{1}_{A}(z, e)f(t_z, t_e)} +
	\rmean{\sum_{(z,t_z) \in Z} \sum_{(e,t_e) \in E}\mathds{1}_{A}(e, z)f(t_e, t_z)}
	\nonumber\\
	&+
	\rmean{\sum_{(e_1,t_{e_1}),(e_2,t_{e_2}) \in E^2}^{\neq}\mathds{1}_{A}(e_1,e_2)f(t_{e_1}, t_{e_2})}. \nonumber
	\end{align}
	Using \Cref{alpha2groundY} and \ref{palm2Y}, recalling that $E$ is a Poisson process independent of $Z$, and using that $f$ is symmetrical, we see that
	\begin{align}
	\alpha_{f}^{(2)}(A) &= 
	\gamma_1(f) \int \alpha_{\ground Y_{0}}^{(2)}(A-(x,x))\intf{X}dx \\
	&+
	\gamma_2(f) \int (\mint{\ground Y_{0}})^2(A-(x_1,x_2)) d\alpha_{\ground X}^{(2)}(x_1,x_2) \nonumber\\
	&+
	2\gamma_2^{EZ}(f) \int_A \intf{Z} \intf{E} d(o_1,o_2)
	\nonumber\\
	&+
	\gamma_2^E(f)\int_A \intf{E}^2 d(o_1,o_2), \nonumber\\
	&=
	\gamma_1(f) \rmean{G(G-1)}\intf{X} \int_A \int \dispdense(o_1-x)\dispdense(o_2-x)dx d(o_1,o_2)\\
	&+
	\gamma_2(f)\intf{X}^2\rmean{G}^2 \int_{A} \int \pcf{X}(||x_1-x_2||) h(o_1-x_1)h(o_2-x_2)  d(x_1,x_2) d(o_1,o_2) \nonumber\\
	&+
	2\gamma_2^{EZ}(f) \intf{Z} \intf{E} \int_A d(o_1,o_2)
	\nonumber\\
	&+
	\gamma_2^E(f)\intf{E}^2 \int_A d(o_1,o_2). \nonumber
	\end{align}
	Write $m$ for the Lebesgue measure on $\R^d$. Then, using the rotational symmetry of $\dispdense$ and $\pcf{X}$, it follows that $\frac{\partial\alpha_{f}^{(2)}}{\partial m^2}(o_1,o_2)$ depends only on $r = ||o_1-o_2||$, and
	\begin{equation}
	\frac{\partial\alpha_{f}^{(2)}}{\partial m^2}(r) = \gamma_1(f) n_c\intf{Z} (\dispdense*\dispdense)(r) + \gamma_2(f)\intf{Z}^2 (\dispdense*\pcf{X})(r) + 2\gamma_2^{EZ}(f) \intf{Z} \intf{E}+\gamma_2^E(f)\intf{E}^2,
	\end{equation}
	where
	\begin{equation}
	(\dispdense*\pcf{X})(r) = \int \pcf{X}(||x_1-x_2||) h(o_1-x_1)h(o_2-x_2)  d(x_1,x_2),
	\end{equation}
	for $||o_1-o_2|| = r$, and in particular
	\begin{align}
	\gamma_2^O(f)\pcf{O}\mcor{O}{f}(r) &= \intf{O}^{-2}\frac{\partial\alpha_{f}^{(2)}}{\partial m^2}(r) \\
	&= 
	\gamma_1(f) \frac{\eta}{\intf{O}}n_c (\dispdense*\dispdense)(r) + \gamma_2(f)\eta^2 ((\dispdense*\pcf{X})(r)-1) + \gamma_2^O(f).
	\end{align}
	By setting $f = 1$ we see that
	\begin{equation}
	\pcf{O} =  \frac{\eta}{\intf{O}}n_c (\dispdense*\dispdense)(r) + \eta^2 ((\dispdense*\pcf{X})(r)-1) + 1,
	\end{equation}
	and using this above we obtain the desired equation
	\begin{equation} 
	\gamma_2^O(f)\mcor{O}{f}(r)\pcf{O}(r) = (\depmark{f}-\indmark{f})\left[\frac{\sigfrac}{\intf{O}}n_c(\dispdense*\dispdense)(r)\right] +\indmark{f}\left[\pcf{O}(r)-1\right]+\gamma_2^O(f).
	\end{equation}

	\begin{center}
		SECTION B: ESTIMATION PROCEDURES IN THE PALM-IBCPP
	\end{center}
	\renewcommand{\thesection}{B.1}
	\section{Extracting spatially invariant statistics from data}
	In this section we will consider how to extract estimators for the quantities
	\begin{equation*}
	\zeta_u = \left(\gamma_1(f_u)-\gamma_2(f_u)\right)n_c, \quad u\in T.
	\end{equation*}
	Recall that for an IBCpp, we have
	\begin{equation} \label{zeta}
	\zeta_u = \frac{\left(\gamma_2^O(f_u)\mcor{O}{f_u}(r)\pcf{O}(r)-\indmark{f_u}\left[\pcf{O}(r)-1\right]-\gamma_2^O(f_u)\right)\intf{O}}{(\dispdense*\dispdense)(r)\sigfrac}, \quad  u\in T,
	\end{equation}
	and estimation of $\zeta_u$ is thus naturally done via estimators for each component on the right hand side. Starting with $\eta$ and $\intf{O}$, these are both functions of the spatial intensities of $O$ and $E$. The standard estimator for the spatial intensity of a point process is the relative number of points per area. In particular, if we have access to $O$ and $E$ in separate windows, we set
	\begin{align}
	\eintf{E} &= \frac{N_E}{|W_E|}, \\
	\eintf{O} &= \frac{N}{|W|},
	\end{align}
	where e.g. $|W|$ is the area of $W$, and consequently we get
	\begin{align}
	\hat{\eta} = 1-\frac{\eintf{E}}{\eintf{O}}.
	\end{align}
	If we do not have access to $E$ in this way, or if we do not wish to account for background noise, we set instead $\hat{\eta} = 1$. 
	
	Moving on to estimators for the pair- and mark correlation functions, $\epcf{O}(r)$ and $\emcor{O}{f_u}(r)$, these are easily obtained using a number of standard implementations, for instance using the kernel smoothing estimators in the R package Spatstat, or by numerical differentiation of the mark-weighted K function \citep[p.~646]{AdrianEgeRolf}, which is significantly faster for large datasets, and is the method used in the supplied code. One detail that must be dealt with, however, is which spatial distances, $r$, we wish to consider. A default choice that emphasizes distances reflecting the spatial scale of blinking clusters is suggested in Algorithm 1 of the main text. 
	
	Next, for the cluster autoconvolution $(\dispdense*\dispdense)$, note first that the density of $P_{\epsilon}$, $\dispdense$, can be obtained as a mean over Gaussian densities where the variance follows $P_{\sigma}$. By changing the order of mean and the integration we thus obtain
	\begin{equation}
	(\dispdense*\dispdense)(r) = \int \dispdense(y_1-x)\dispdense(y_2-x) dx = \rmean{\frac{e^{-\frac{r^2}{2(\sigma_1^2+\sigma_2^2)}}}{2\pi (\sigma_1^2+\sigma_2^2)}},
	\end{equation}
	where the mean is with respect to $\sigma_1$ and $\sigma_2$ independently following $P_{\sigma}$. We do not know $P_{\sigma}$, but we do have predictions of $\sigma_k$ in $\hat{\sigma}_k$ for each $k \in \{1,2,..,N\}$, and the natural estimator of $(\dispdense*\dispdense)$ is then to replace $P_{\sigma}$ with the empirical distribution, $P_{\hat{\sigma}}$, of the observed localization uncertainties, that is
	\begin{equation}
	\widehat{(\dispdense*\dispdense)}(r) = \rmean{\frac{e^{-\frac{r^2}{2(\hat{\sigma}_1^2+\hat{\sigma}_2^2)}}}{2\pi (\hat{\sigma}_1^2+\hat{\sigma}_2^2)}}, \quad r \in R.
	\end{equation}
	with $\hat{\sigma}_1$ and $\hat{\sigma}_2$ independently following $P_{\hat{\sigma}}$. This mean can be computed e.g. via sampling $\hat{\sigma}_1$ and $\hat{\sigma}_2$ a larger number of times with replacement from $\{\hat{\sigma}_k\}_{k = 1}^N$. 
	
	Finally, we need estimators of $\gamma_2(f_u)$ and $\gamma_2^O(f_u)$. Here, $\gamma_2^O$ is more well-known as the normalization constant in the mark correlation function, and a standard estimator is 
	\begin{equation}
	\hat{\gamma}_2^O(f_u) = \frac{1}{N(N-1)}\sum_{i,j}^{\neq} \mathds{1}(|t_{o_i}-t_{o_j}| \le u), 
	\end{equation}
	see \citep[p.~393]{Gelfand2010}. As the number of pairs in this sum can be quite large, a less computationally expensive estimator first sub-samples a smaller number of pairs to sum over. Next, for $\gamma_2(f_u)$, note first that $\gamma_2^O(f_u)$ has a more formal description as the mean 
	\begin{equation}
	\gamma_2^O(f_u) = \int_0^{\infty}\int_0^{\infty} f(t_1,t_2) d\palms{1}{O}(t_1)d\palms{1}{O}(t_2),
	\end{equation}
	where $\palms{1}{O}$ is the 1-point mark distribution of $O$. This is important in the context of estimating $\gamma_2(f_u)$ since we have similarly
	\begin{equation}
	\gamma_2(f_u) = \int_0^{\infty}\int_0^{\infty} f(t_1,t_2) d\palms{1}{Z}(t_1)d\palms{1}{Z}(t_2),
	\end{equation}
	where $\palms{1}{Z}$ is the 1-point mark distribution of the blinking clusters in $Z$, which is connected to $\palms{1}{O}$ by the identity
	\begin{equation}
	\palms{1}{O}(t) = \sigfrac \palms{1}{Z}(t)+(1-\sigfrac)\frac{t}{b},
	\end{equation}
	c.f. \hyperref[suppA]{Section A}. This suggests that we can estimate $\gamma_2(f_u)$ by first estimating the $\palms{1}{Z}$ using the empirical mark distribution $\hat{M}_O^{(1)}$ via
	\begin{equation} \label{markzdist}
	\hat{M}_Z^{(1)}(t) = \frac{\hat{M}_O^{(1)}(t)-(1-\hat{\eta})\frac{t}{b} }{\hat{\eta}}
	\end{equation}
	and finally computing
	\begin{equation}
	\hat{\gamma}_2(f_u) = \rmean{\mathds{1}(|t_1-t_2|)},
	\end{equation}
	where $t_1$ and $t_2$ follow $\hat{M}_Z^{(1)}$. This can be done by sampling from $\hat{M}_Z^{(1)}$ a large number of times, which can be accomplished using for instance the method of inverse transform sampling. 
	
	We are finally in a position to extract $\zeta_u$. Since \Cref{zeta} states that the denominator and enumerator on the right hand side are proportional for each $r$, a least squares fit suggests the estimators
	\begin{equation}
	\hat{\zeta}_u = \frac{\eintf{O}}{\hat{\sigfrac}}\frac{\sum_{r \in R}\left[\hat{\gamma}_2^O(f_u)\emcor{O}{f_u}(r)\epcf{O}(r)-\hat{\gamma}_2(f_u)(\epcf{O}(r)-1)-\hat{\gamma}_2^O(f_u)\right]\left[\widehat{(\dispdense*\dispdense)}(r)\right]}{\sum_{r \in R} \left[\widehat{(\dispdense*\dispdense)}(r)\right]^2},
	\end{equation} 
	for each $u\in T$. 
	
	\renewcommand{\thesection}{B.2}
	\section{Kinetic rate estimation}
	With the spatially invariant statistics in $\{\hat{\zeta}_u\}_{u \in T}$ at hand, we are able to estimate the kinetic rates. We set up the weighted minimum contrast problem
	\begin{equation} \label{minprob}
	\min_{\hat{r}_D,\hat{r}_R,\hat{r}_B} \sum_{u \in T}\sum_{r \in R} \left(\frac{\hat{\zeta}_u}{\hat{\gamma}_2(f_u)}\right)^2\left(\hat{\zeta}_u-(\gamma_1(f_u)-\hat{\gamma}_2(f_u))n_c\right)^2,
	\end{equation}
	where $\frac{\hat{\zeta}_u}{\hat{\gamma}_2(f_u)}$ are weights chosen to emphasize the $\zeta_u$ that are most informative. These weights are motivated by the fact that
	\begin{equation*}
	\frac{\zeta_u}{\gamma_2(f_u)} = \left(\frac{\gamma_1(f_u)}{\gamma_2(f_u)}-1\right)n_c,
	\end{equation*}
	puts most weight on $u \in T$ where $\gamma_1(f_u)$ moves between $0$ and $1$, while down-weighing large $u$ for which $\gamma_1(f_u)$ is constantly $1$ and weakly informative. In order to solve the minimization problem in \ref{minprob}, we need to know how $\gamma_1(f_u)$ and $n_c$ depend on $(r_D, r_R, r_B)$, which leads to some rather gritty computations. In fact, we must be satisfied with asymptotically ($\Delta \rightarrow 0$) exact approximations, derivations of which can be found in \hyperref[Appendix:C1]{Section C}. Define the following random variables and associated characteristic functions
	
	\begin{align*}
	&N_b \sim \mathrm{Geom}_1(p), \\
	&W_F \sim \mathrm{Exp}(r_D+r_B), \\
	&W_D \sim \mathrm{Exp}(r_R), \\
	&W_I \sim \mathrm{Exp}(r_F), \\
	&\phi_R(v) = \rmean{e^{ivW_R}}, \\
	&\phi_F(v) = \rmean{e^{ivW_F}}, \\
	&\phi_{(F+R)}(v) = \rmean{e^{ivW_F}}\rmean{e^{ivW_R}},
	\end{align*}
	where $p = \frac{r_B}{r_D+r_B}$ is the \textsl{bleaching probability}, and $\mathrm{Geom}_1$ is a Geometric distribution starting from $1$. Here, $N_b$ has the interpretation as the number of blinks (F-state visits), and $W_F$ is the holding time in state $F$, and similarly for $W_D$ and $W_I$. Next, define the following quantities
	\begin{align*}
	&A(v) = \frac{2\rmean{N_b}
		\left(
		\phi_F(v)e^{-iv\Delta\frac{1}{2}}+\left(\frac{\rmean{W_F}}{\Delta}-\frac{1}{2}\right)(e^{-\Delta iv}-1)-1
		\right)}{(1-e^{-\Delta iv})^2}, \\
	&B(v) = \phi_R(v)\left(
	\rmean{\phi_{(F+R)}(v)^{N_b}}-1-\rmean{N_b}(\phi_{(F+R)}(v)-1)
	\right),  \\
	&C(v) = \frac{2e^{-iv\Delta 2}}{(1-e^{-\Delta iv})^2} \left(\frac{\phi_F(v)e^{iv\Delta\frac{1}{2}}-1}{\phi_{(F+R)}(v)-1}\right)^2, \\
	&D = \rmean{N_b^2}\left( \frac{\rmean{W_F}}{\Delta}+\frac{1}{2}\right)^2 + 
	\rmean{N_b}\left[ \frac{\rmean{W_F^2}-\rmean{W_F}^2}{\Delta^2}-\frac{\rmean{W_F}}{\Delta}-\frac{1}{2} \right],
	\end{align*}
	and the CDF $u \mapsto \gamma_1(f_u)$ then has characteristic function given as approximately
	\begin{equation}
	\phi(v) \approx \frac{A(v) + B(v)C(v)}{D}.
	\end{equation}
	All the involved mean values are elementary to compute, and we can thus obtain our approximate $\gamma_1(f_u)$ by numerically inverting $\phi(v)$, which can be done efficiently using the fast Fourier transform, see e.g. \citep{Hurlimann2013}. 
	
	For $n_c$, we recall that
	\begin{equation}
	n_c  = \frac{\rmean{G^2}}{\rmean{G}}-1,
	\end{equation}
	and we simply plug in the approximations
	\begin{align*}
	\rmean{G} &\approx \rmean{N_b}\left(\frac{\rmean{W_F}}{\Delta}+1\right)- \rmean{N_b-1}\mu_R^1, \\
	\rmean{G^2} &\approx \rmean{N^2_b}\left(\frac{\rmean{W_F}}{\Delta}+1\right)^2+\rmean{N_b}\frac{\rmean{W_F^2}-\rmean{W_F}^2}{\Delta^2} \\
	&+ \rmean{(N_b-1)^2}(\mu_R^1)^2+\rmean{N_b-1}\left(
	\mu_R^2
	-
	(\mu_R^1)^2\right) \\
	&- 2\rmean{N_b(N_b-1)}\left(\frac{\rmean{W_F}}{\Delta}+1\right)\mu_R^1,
	\end{align*}
	with 
	\begin{align*}
	\mu_R^1 &= \frac{r_R\Delta+e^{-r_R\Delta}-1}{r_R\Delta}, \\
	\mu_R^2 &= \frac{2(1-e^{-r_R\Delta}-r_R\Delta)+(r_R\Delta)^2}{(r_R\Delta)^2}.
	\end{align*} 
	
	The functions $f_u$ were selected precisely to eliminate the influence of $r_F$, and $r_F$ consequently plays no role in the minimization problem above. In order to estimate $r_F$ we thus need an additional step. We have the following asymptotically exact relation
	\begin{equation*}
	r_R \approx \left(\frac{1}{2}\gamma_2(f_+) - A_2-B_2\right)^{-1},
	\end{equation*}
	where $f_+(t_1,t_2) = t_1+t_2$, and 
	\begin{align*}
	A_2 &= \frac{\frac{\rmean{W_F^2}}{2\Delta}+\rmean{W_F}+\frac{3\Delta}{8}}{\frac{\rmean{W_F}}{\Delta}+\frac{1}{2}},\\
	B_2 &= \frac{ (\frac{\rmean{W_F}}{\Delta}+\frac{1}{2})(\frac{1}{2}\rmean{N_b(N_b-1)}(\rmean{W_F}+\rmean{W_R})+\rmean{N_b}\Delta\frac{1}{2}) }{\rmean{N_b}(\frac{\rmean{W_F}}{\Delta}+\frac{1}{2})}.
	\end{align*}
	Write $\hat{A}_2$ and $\hat{B}_2$ for $A_2$ and $B_2$ computed with the estimated $(\hat{r}_D, \hat{r}_R, \hat{r}_B)$ in the previous step. We estimate $\gamma_2(f_+)$ directly from the observed timepoints using (\ref{markzdist}), and obtain an estimator for $r_F$ as
	\begin{equation}
	\hat{r}_F = \left(\frac{\frac{1}{N}\sum_{i = 1}^{N}t_{o_i}-(1-\hat{\eta})\frac{b}{2} }{\hat{\eta}} -\hat{A}_2-\hat{B}_2\right)^{-1}.
	\end{equation}
	If the dataset recording was stopped too early, $\hat{r}_F$ may be subject to censoring biases, as we then only observed blinking clusters beginning before time $b$, and $\hat{r}^{-1}_F$ is then rather estimating the mean of the conditional distribution $(W_I | W_I < b)$. A corrected estimate can be found by equating this mean with its theoretical counterpart, i.e. solving
	\begin{equation*}
	\frac{e^{r^c_F b}-r^c_F b-1}{r^c_F(e^{r^c_Fb}-1)}-\hat{r}_F^{-1} = 0,
	\end{equation*}
	in $r^c_F$.

	\vspace{1cm}
	\begin{center}
		SECTION C: APPROXIMATE DISCRETIZED STATISTICS
	\end{center}
	
	\renewcommand{\thesection}{C.1}
	\section{Approximate \texorpdfstring{$\phi(v)$}{phi(v)}} \label{Appendix:C1}
	The mean value to compute is formally
	\begin{equation}
	\phi(v) := 
	\frac{ 
		\rmean{
			\sum_{j_1,j_2 \in \{1,..,G\}}^{\neq} e^{iv|m_{j_1}-m_{j_2}|}
	}}{\rmean{G(G-1)}},
	\end{equation}
	where we have dropped the heavier notation of timepoints in the main text, so that $(m_{j_1},m_{j_2})$ are arrival times (marks) $j_1$ and $j_2$ in the typical blinking cluster. Denote again by $N_b$ the number of $F$-state visits (number of blinks), and by $F_s$ the observed timepoints between the entrance to the $s$'th and $(s+1)$'th $F$-state visits for $s < N_b$, and $F_{N_b}$ are all observed timepoints after the last entrance to the $F$-state. Below, we will assume w.l.o.g. that the timepoints are sorted, that is  $m_{j_1} < m_{j_2}$ for $(j_2 > j_1)$ when $(m_{j_1}, m_{j_2}) \in F_s$ - this is entirely as a notational convenience. We can split the summation according to whether $m_{j_1}$ and $m_{j_2}$ are from the same $F_s$, and otherwise how many $F$-state visits are separating them. Thus
	\begin{align}
	\phi(v) &= 
	\frac{ 
		\rmean{\sum_{s = 1}^{N_b}\sum_{(m_{j_1},m_{j_2}) \in F_s}^{\neq} e^{iv|m_{j_1}-m_{j_2}|}}
	}{
		\rmean{G(G-1)}
	}
	\\&+
	\frac{\rmean{ 
			\sum_{s_1 = 1}^{N_b}\sum_{s_2 = 1}^{N_b}\mathds{1}(s_1 \neq s_2)\sum_{m_{j_1} \in F_{s_1}}\sum_{m_{j_2} \in F_{s_2}} e^{iv|m_{j_1}-m_{j_2}|}
	}}{
		\rmean{G(G-1)}
	}. \nonumber
	\end{align}
	To compute these terms, referred to as the ''non-separated'' and ''separated'' terms, respectively, we write the involved quantities in terms of a continuous part, and an error part, and demonstrate that the errors vanish asymptotically, and in particular can be ignored for a given framerate as a valid approximation. 
	
	First, we consider the number of timepoints in $F_s$, $|F_s|$. Since only those frames that do not fully overlap the signal from the $s$'th $F$-visit (of which there are at most 2) cause discretization effects, we can write
	\begin{equation} \label{fs_exp}
	|F_s| = \frac{W_{F_s}}{\Delta} + E^F_s,
	\end{equation}
	where $W_{F_s}$ is the waiting time that was spent on the $s$'th visit to the $F$ state, and $E^F_s$ is an error term with
	\begin{equation}
	P(E^F_s \in  (-1,2)) = 1,
	\end{equation}
	and in particular we obtain for $G$
	\begin{equation}
	G = \sum_{s = 1}^{N_b} |F_{s}| = \sum_{s = 1}^{N_b} \frac{W_{F_s}}{\Delta} + \sum_{s = 1}^{N_b} E^F_s.
	\end{equation}
	Next, consider the inner sum from the non-separated term:
	\begin{equation}
	\sum_{(m_{j_1},m_{j_2}) \in F_s}^{\neq} e^{iv|m_{j_1}-m_{j_2}|}.
	\end{equation}
	Here, note that the first observed timepoint in $F_s$, $m^s_1$, can be written as
	\begin{equation}
	m^s_1 = E^{m_1}_s + W_I+\sum_{k = 1}^{s-1} (W_{F_k}+W_{R_k}),
	\end{equation}
	since there is always a waiting time of $W_I$ spent in the inactive state, and $(s-1)$ visits in and out of the $F$ state before the $s$'th visit. $E^{m_1}_s$ is again a discretization error, with magnitude
	\begin{equation}
	P(E_s^{m_1} \in (0, \Delta)) = 1.
	\end{equation}
	Since each member of $F_s$ is a whole number of $\Delta$-increments away from $m_1^s$, this in particular means that, for $(m_{j_1},m_{j_2}) \in F_s$ with $j_2 > j_1$:
	\begin{equation}
	|m_{j_1}-m_{j_2}| = (j_2-j_1)\Delta,
	\end{equation}
	and any discretization effects, and the time spent in the $I$-state, can be seen to disappear here. We can now expand on the non-separate term enumerator:
	\begin{align}
	&\rmean{\sum_{s = 1}^{N_b}\sum_{(m_{j_1},m_{j_2}) \in F_s}^{\neq} e^{iv|m_{j_1}-m_{j_2}|}}
	\\ =
	&\rmean{\sum_{s = 1}^{N_b} \sum_{(j_1,j_2) \in \{1,2,..,|F_s|\}}^{\neq} e^{iv(j_1 \vee j_2 - j_1 \wedge j_2)\Delta}} \\=
	&2\rmean{\sum_{s = 1}^{N_b} \sum_{j = 1}^{|F_s|-1} (|F_s|-j)e^{iv j\Delta}} \\=
	&2\rmean{\sum_{s = 1}^{N_b} \frac{e^{iv\Delta(|F_s|-1)}+e^{-iv\Delta}(|F_s|-1)-|F_s|}{(e^{-iv\Delta}-1)^2}} \\=
	&2\rmean{\sum_{s = 1}^{N_b} 
		\frac{e^{ivW_{F_s}}e^{iv\Delta(E^F_s-1)}+e^{-iv\Delta}(\frac{W_{F_s}}{\Delta}+E^F_s-1)-\frac{W_{F_s}}{\Delta}-E^F_s}{(e^{-iv\Delta}-1)^2}},
	\end{align}
	At this point, consider what happens in the limit as $\Delta \rightarrow 0$ for the complete non-separate term:
	\begin{align}
	&\lim_{\Delta \rightarrow 0} \frac{\rmean{\sum_{s = 1}^{N_b}\sum_{(m_{j_1},m_{j_2}) \in F_s}^{\neq} e^{iv|m_{j_1}-m_{j_2}|}}}{\rmean{G(G-1)}} \\=
	&\frac{2\rmean{\lim_{\Delta \rightarrow 0}\sum_{s = 1}^{N_b} 
			e^{ivW_{F_s}}e^{iv\Delta(E^F_s-1)} + \frac{W_{F_s}}{\Delta}(e^{-iv\Delta}-1)+E^F_s(e^{-iv\Delta}-1)-e^{-iv\Delta} }}{\lim_{\Delta \rightarrow 0}(e^{-iv\Delta}-1)^2\left[\rmean{\left(\sum_{s = 1}^{N_b} \frac{W_{F_s}}{\Delta} + \sum_{s = 1}^{N_b} E^F_s\right)^2}-\rmean{\sum_{s = 1}^{N_b} \frac{W_{F_s}}{\Delta} + \sum_{s = 1}^{N_b} E^F_s}\right]} \\=
	&\frac{2\rmean{\sum_{s = 1}^{N_b} 
			1+ivW_{F_s}-e^{iW_{F_s}}}}{v^2\rmean{\left(\sum_{s = 1}^{N_b} W_{F_s}\right)^2}}
	\\=
	&\frac{2\rmean{N_b}(1+iv\rmean{W_{F}}-\phi_F(u))}{v^2(\rmean{N_b}\rmean{W_F^2}+\rmean{N_b(N_b-1)}\rmean{W_F}^2)}.
	\end{align}
	Predictably the rounding errors play no role in the limit, and as a simple approximation we therefore set $E_s^F = \frac{1}{2}$ to the midpoint of its domain for all $s$, to obtain the asymptotically exact approximation:
	\begin{align}
	&\frac{\rmean{\sum_{s = 1}^{N_b}\sum_{(m_{j_1},m_{j_2}) \in F_s}^{\neq} e^{iv|m_{j_1}-m_{j_2}|}}}{\rmean{G(G-1)}} \\ \approx 
	&\frac{2\rmean{\sum_{s = 1}^{N_b} 
			e^{ivW_{F_s}}e^{-iv\Delta\frac{1}{2}}+e^{-iv\Delta}(\frac{W_{F_s}}{\Delta}-\frac{1}{2})-\frac{W_{F_s}}{\Delta}-\frac{1}{2} }
	}{(e^{-iv\Delta}-1)^2\left[\rmean{\left(\sum_{s = 1}^{N_b} \frac{W_{F_s}}{\Delta}+\frac{1}{2} \right)^2}-\rmean{\sum_{s = 1}^{N_b} \frac{W_{F_s}}{\Delta}+\frac{1}{2}}\right]} \\ =
	&\frac{2\rmean{N_b}(\phi_F(v)e^{-iv\frac{\Delta}{2}}
		+e^{-iv\Delta}(\frac{\rmean{W_F}}{\Delta}-\frac{1}{2})-\frac{\rmean{W_F}}{\Delta}-\frac{1}{2}
		)}{(e^{-iv\Delta}-1)^2 \left(\rmean{N_b^2}\left( \frac{\rmean{W_F}}{\Delta}+\frac{1}{2} \right)^2 + 
		\rmean{N_b}\left[ \frac{\rmean{W_F^2}-\rmean{W_F}^2}{\Delta^2}-\frac{\rmean{W_F}}{\Delta}-\frac{1}{2} \right]\right)},
	\end{align}
	which is $\frac{A(v)}{D}$. 
	
	Now, consider the separate summation enumerator. We use similar techniques as before. Note that for $F_{s_1}$ and $F_{s_2}$ there are $|s_1-s_2-1|$ $W_F$ waiting times, and $|s_1-s_2|$ $W_R$ waiting times, separating the closest pair in $F_{s_1} \times F_{s_2}$, up to rounding error. Thus, if we enumerate the timepoints in $F_{s_1}$ instead starting from the end (so that $m'_j \in F_{s_1}$ is the $j$'th largest value in $F_s$, $j \ge 1$), the differences in timepoints $m'_{j_1} \in F_{s_1}$ and $m_{j_2} \in F_{s_2}$, with $s_2 > s_1$, can be written on the form. 
	\begin{equation}
	|m'_{j_1}-m_{j_2}| = W_{s_2} + \sum_{k = 1}^{s_2-s_1}(W_{R_{s_1+k}}+W_{F_{s_1+k}})+ (j_1+j_2-2)\Delta + E_{(s_1, s_2)},
	\end{equation}
	where $E_{(s_1, s_2)}$ only depends on $(s_1, s_2)$ and has 
	\begin{align}
	P(E_{(s_1, s_2)} \in (-\Delta,\Delta)) = 1.
	\end{align}
	Therefore:
	\begin{align}
	&\sum_{s_1 = 1}^{N_b}\sum_{s_2 = 1}^{N_b}\mathds{1}(s_1 \neq s_2)\sum_{m_{j_1} \in F_{s_1}}\sum_{m_{j_2} \in F_{s_2}} e^{iv|m_{j_1}-m_{j_2}|} \\&=
	2\sum_{s_1 = 1}^{N_b-1}\sum_{s_2 = s_1+1}^{N_b}\sum_{m_{j_1} \in F_{s_1}}\sum_{m_{j_2} \in F_{s_2}} e^{iv|m_{j_1}-m_{j_2}|} \\&=
	2\sum_{s_1 = 1}^{N_b-1}\sum_{s_2 = s_1+1}^{N_b} e^{iv W_{s_2}}e^{iv\sum_{k = 1}^{s_2-s_1}(W_{R_{s_1+k}}+W_{F_{s_1+k}})}e^{ivE_{(s_1, s_2)}}\\&\times \nonumber e^{-iv\Delta 2}\sum_{j_1 = 1}^{|F_{s_1}|}\sum_{j_2 = 1}^{|F_{s_2}|} e^{iv(j_1+j_2)\Delta} \\&=
	2\sum_{s_1 = 1}^{N_b-1}\sum_{s_2 = s_1+1}^{N_b} e^{iv W_{s_2}}e^{iv\sum_{k = 1}^{s_2-s_1}(W_{R_{s_1+k}}+W_{F_{s_1+k}})}e^{ivE_{(s_1, s_2)}}\\&\times \nonumber e^{-iv\Delta 2}\frac{(e^{iv\Delta |F_{s_1}|}-1)(e^{iv\Delta |F_{s_2}|}-1)}{(e^{iv\Delta}-1)^2}.
	\end{align}
	At this point, it should be clear that discretization effects again have no impact in the limit. For the sake of completion, we compute also this asymptotic value:
	\begin{align}
	&\lim_{\Delta \rightarrow 0} \frac{\rmean{\sum_{s_1 = 1}^{N_b}\sum_{s_2 = 1}^{N_b}\mathds{1}(s_1 \neq s_2)\sum_{m_{j_1} \in F_{s_1}}\sum_{m_{j_2} \in F_{s_2}} e^{iv|m_{j_1}-m_{j_2}|}}}{\rmean{G(G-1)}} \\&=
	\frac{-\rmean{2\sum_{s_1 = 1}^{N_b-1}\sum_{s_2 = s_1+1}^{N_b} e^{iv W_{s_2}}e^{iv\sum_{k = 1}^{s_2-s_1}(W_{R_{s_1+k}}+W_{F_{s_1+k}})}(e^{ivW_{F_{s_1}}}-1)(e^{iv W_{F_{s_2}}}-1)}}{v^2(\rmean{N_b}\rmean{W_F^2}+\rmean{N_b(N_b-1)}\rmean{W_F}^2)} \\&=
	\frac{     
		2\left(\frac{\phi_F(v)-1}{\phi_{(F+R)}(v)-1}\right)^2
		\phi_R(v)\left(
		1+\rmean{N_b}(\phi_{(F+R)}(v)-1)-\rmean{\phi_{(F+R)}(v)^{N_b}}
		\right)
	}{v^2(\rmean{N_b}\rmean{W_F^2}+\rmean{N_b(N_b-1)}\rmean{W_F}^2)}.
	\end{align}
	Thus, replacing again all discretization errors with the midpoints of their domains ($E^F_s = \frac{1}{2}, E_{s_1,s_2} =  0$), we get an asymptotically exact approximation:
	\begin{align}
	&\frac{\rmean{\sum_{s_1 = 1}^{N_b}\sum_{s_2 = 1}^{N_b}\mathds{1}(s_1 \neq s_2)\sum_{m_{j_1} \in F_{s_1}}\sum_{m_{j_2} \in F_{s_2}} e^{iv|m_{j_1}-m_{j_2}|}}}{\rmean{G(G-1)}} \\ \approx
	&\frac{     
		2e^{-iv\Delta 2}\left(\frac{\phi_F(v)e^{iv\Delta\frac{1}{2}}-1}{\phi_{(F+R)}(v)-1}\right)^2
		\phi_R(v)\left(
		\rmean{\phi_{(F+R)}(v)^{N_b}}-1-\rmean{N_b}(\phi_{(F+R)}(v)-1)
		\right)
	}{
		(e^{-iv\Delta}-1)^2 \left(\rmean{N_b^2}\left( \frac{\rmean{W_F}}{\Delta}+\frac{1}{2} \right)^2 + 
		\rmean{N_b}\left[ \frac{\rmean{W_F^2}-\rmean{W_F}^2}{\Delta^2}-\frac{\rmean{W_F}}{\Delta}-\frac{1}{2} \right]\right)
	},
	\end{align}
	which is $\frac{B(v)C(v)}{D}$.

	\renewcommand{\thesection}{C.2}
	\section{Approximate \texorpdfstring{$n_c$}{nc}}
	We wish to compute
	\begin{equation}
	n_c = \frac{\rmean{G^2}}{\rmean{G}}-1.
	\end{equation}
	Instead of approximating the moments directly, we first approximate the distribution of $G$, from which the moments can be obtained. We can write somewhat loosely
	\begin{equation}
	G = \sum_{s = 1}^{N_b} \#(\text{frames hit by the s'th F-signal}) - \sum_{s = 1}^{N_b-1} \mathds{1}(\text{F-signals s and s+1 share a frame}),
	\end{equation}
	where by "sharing" we mean that the continuous time signals emitted from the 2 $F$-state visits hit the same frame. Now, computing the distribution of $G$ from this representation is made intractable due to the dependence and complicated behavior in the summands caused by disretization to the fixed grid $\Delta \Z $. Instead, we replace the summands with their mean under disretization to grids $\Delta \Z +U$, where $U \sim Uni(0,\Delta)$. Write $\smean{U}{\cdot}$ for this mean, and let $\floor{a}$ and $\frap{a}$ denote the integer and fractional parts, respectively, of a number $a$. Write $T^I_s$ and $T^O_s$ for the entrance and exit times, respectively, for the $s$'th $F$-state visit, and $D_s$ for the distance from $T^I_s$ to the nearest gridpoint larger than $T^I_s$. Then, we obtain for any $s$
	\begin{align}
	&\smean{U}{\#(\text{frames hit by the s'th F-signal})} \\&=
	\floor{W_{F_s}\Delta^{-1}} + \smean{U}{2\mathds{1}(D_s < \frap{W_{F_s}\Delta^{-1}})+\mathds{1}(D_s > \frap{W_{F_s}\Delta^{-1} })} \\&=
	\floor{W_{F_s}\Delta^{-1}} + 2\frap{W_{F_s}\Delta^{-1}}+(1-\frap{W_{F_s}\Delta^{-1} }) \\&=
	W_{F_s}\Delta^{-1}+1.
	\end{align}
	Now, for the second sum, we get
	\begin{align}
	&\smean{U}{\mathds{1}(\text{F-signals s and s+1 share a frame})} \\&=
	1-\smean{U}{\mathds{1}(\text{there is a gridpoint between $T^O_{s}$ and $T^I_{s+1}$ } ) } \\ &=
	1-(W_{R_s}\Delta^{-1}\mathds{1}(W_{R_s} \le \Delta)+\mathds{1}(W_{R_s} > \Delta)) \\ &=
	\mathds{1}(W_{R_s}\Delta^{-1} \le 1)(1-W_{R_s}\Delta^{-1}),
	\end{align}
	and our approximation for $G$ is thus
	\begin{equation}
	G \approx \sum_{s = 1}^{N_b} \frac{W_{F_s}}{\Delta}+1 - \sum_{s = 1}^{N_b-1} \mathds{1}(\frac{W_{R_s}}{\Delta} \le 1)(1-\frac{W_{R_s}}{\Delta}),
	\end{equation}
	from which we obtain
	\begin{equation}
	\rmean{G} \approx \rmean{N_b}\left(\frac{\rmean{W_F}}{\Delta}+1\right)- \rmean{N_b-1}\mu_R^1,
	\end{equation}
	where $\mu_R^1 = \int_{0}^{1}(1-x)dP_{\frac{W_R}{\Delta}}(x)$, and
	\begin{align}
	\rmean{G^2} &\approx \rmean{N^2_b}\left(\frac{\rmean{W_F}}{\Delta}+1\right)^2+\rmean{N_b}\frac{\rmean{W_F^2}-\rmean{W_F}^2}{\Delta^2} \\
	&+ \rmean{(N_b-1)^2}(\mu_R^1)^2+\rmean{N_b-1}\left(
	\mu_R^2
	-
	(\mu_R^1)^2\right) \nonumber\\
	&- 2\rmean{N_b(N_b-1)}\left(\frac{\rmean{W_F}}{\Delta}+1\right)\mu_R^1,\nonumber
	\end{align}
	with $\mu_R^2 = \int_{0}^{1}(1-x)^2dP_{\frac{W_R}{\Delta}}(x)$. 
	
	If we write $n_c(\Delta)$ for the approximation of $n_c$ given a framerate of $\Delta^{-1}$, we have that $n_c(\Delta)$ is asymptotically exact in the sense that, after appropriate normalization, we have
	\begin{equation}
	\lim_{\Delta \rightarrow 0} \Delta n_c(\Delta) = \lim_{\Delta \rightarrow 0} \Delta n_c,
	\end{equation}
	where this asymptotic value is given as
	\begin{equation}
	\lim_{\Delta \rightarrow 0} \Delta n_c = \frac{\rmean{N_b}\rmean{W_F^2}+\rmean{N_b(N_b-1)}\rmean{W_F}^2 }{\rmean{N_b}\rmean{W_F}}.
	\end{equation}

	\renewcommand{\thesection}{C.3}
	\section{Approximate \texorpdfstring{$\gamma_2(f_{+})$}{gamma2(d)}} 
	By definition, we have
	\begin{equation}
	\gamma_2(f_{+}) = \frac{\rmean{\sum_{k = 1}^{G}\sum_{j = 1}^{G'} d(m_k,m_j') } }{\rmean{G}^2} = \frac{\rmean{\sum_{k = 1}^{G}\sum_{j = 1}^{G'} m_k+m_j' } }{\rmean{G}^2},
	\end{equation}
	where we again drop the drop the heavier time point notation, such that e.g. $m_k$ is arrival time $k$ in a typical cluster, and $m_j'$ is arrival time $j$ in an independent copy of the typical cluster. Clearly, then,
	\begin{equation}
	\frac{1}{2}\gamma_2(f_{+}) = \frac{\rmean{\sum_{k = 1}^{G}m_k} }{\rmean{G}}.
	\end{equation}
	Now, write $T_s^I$ for the (continuous) entrance time to the $s$'th $F$-state visit. Then the first observed timepoint in $F_s$ can be written as $T_s^I + E_s$, where $E_s$ is a discretization error with $P(0\le E_s \le \Delta) = 1$. Note further, that 
	\begin{equation}
	T_s^I = W_{I} + \sum_{i = 1}^{s-1}(W_{F_i}+W_{R_i}),
	\end{equation}
	and we arrive at the expression
	\begin{align}
	\frac{1}{2}\gamma_2(f_{+}) &= \frac{\rmean{\sum_{s = 1}^{N_b}|F_s|(T_s^I+E_s)+\sum_{k = 1}^{|F_s|}k\Delta} }{\rmean{G}} \\&=
	\frac{\rmean{\sum_{s = 1}^{N_b}|F_s|(T_s^I+E_s)} }{\rmean{G}}+
	\frac{\Delta\rmean{\sum_{s = 1}^{N_b} |F_s|(|F_s|+1)}}{2\rmean{G}}.
	\end{align}
	Now, setting everywhere $|F_s| = \frac{W_{F_s}}{\Delta}+\frac{1}{2}$ as in \hyperref[Appendix:C1]{Section C.1}, and similarly setting all $E_s = \Delta\frac{1}{2}$, we get
	\begin{align}
	\frac{\rmean{\sum_{s = 1}^{N_b}\sum_{k = 1}^{|F_s|}k\Delta} }{\rmean{G}} =  \frac{\rmean{N_b}(\frac{\rmean{W_F^2}}{2\Delta}+\rmean{W_F}+\frac{3\Delta}{8})}{\rmean{G}},
	\end{align}
	and
	\begin{align}
	&\frac{\rmean{\sum_{s = 1}^{N_b}|F_s|(T_s^I+E_s)} }{\rmean{G}} \\&= 
	\rmean{W_I}+\frac{\rmean{\sum_{s = 1}^{N_b}(\frac{W_{F_s}}{\Delta}+\frac{1}{2})(\sum_{i = 1}^{s-1}(W_{F_i}+W_{R_i})+\Delta\frac{1}{2})} }{\rmean{G}} \\&=
	\rmean{W_I}+\frac{(\frac{\rmean{W_F}}{\Delta}+\frac{1}{2})(\frac{1}{2}\rmean{N_b(N_b-1)}(\rmean{W_F}+\rmean{W_R})+\rmean{N_b}\Delta\frac{1}{2}) }{\rmean{G}},
	\end{align}
	so that using $\rmean{G} \approx \rmean{N_b}\left(\frac{\rmean{W_F}}{\Delta}+\frac{1}{2}\right)$
	yields the approximation. Again, the approximation is asymptotically exact, with limiting value
	\begin{equation}
	\lim_{\Delta \rightarrow 0} \frac{1}{2}\gamma_2(f_{+}) = \rmean{W_I}+\frac{\rmean{W_F}(\frac{1}{2}\rmean{N_b(N_b-1)}(\rmean{W_F}+\rmean{W_R}) }{\rmean{N_b}\rmean{W_F}} +
	\frac{\rmean{W_F^2}}{2\rmean{W_F}}.
	\end{equation}

	\vspace{1cm}
	\renewcommand{\thesection}{D}
	\begin{center}
		SECTION D: USE ON GENERAL PROTEIN SAMPLES
	\end{center}
	\section{Use on general general protein samples} \label{Appendix:C:Genx}
	In this section we show that we can use the same estimation procedures from the main text on samples with general distribution for $\ground X$, and still expect meaningful estimates. We assume here that the spatial dimension is $2$, but the same arguments can be made in arbitrary dimension with minor changes. 
	
	Assume that the IBCpp $O$ is observed with $N$ points in $W\times [0,b]$. Standard estimators of $\gamma_2^O(f)\mcor{O}{f}$ and $\pcf{O}$, if $O$ were motion-invariant, are given as
	\begin{align}
	\hat{\gamma}_2^O(f)\emcor{O}{f}(r) &= \frac{\sum_{i \neq j}f(t_{o_i},t_{o_j})\kappa(||o_i-o_j|| - 	r)w(o_i,o_j)\mathds{1}_W(o_i,o_j)}{\sum_{i \neq j} \kappa(||o_i-o_j|| - r)w(o_i,o_j)\mathds{1}_W(o_i,o_j)}, \\
	\epcf{O}(r)					       &= c(r)\sum_{i \neq j}\kappa(||o_i-o_j|| - r)w(o_i,o_j)\mathds{1}_W(o_i,o_j).
	\end{align}
	Here, $c(r) = (2\pi r)^{-1}N^{-2}|W|$, $\kappa$ is a smoothing kernel, $w(x,y)$ are edge correction weights, and $\mathds{1}_W(o_i,o_j)$ is the indicator that both $o_i$ and $o_j$ are in the set $W$, see e.g. \citep[p.~308, 393]{Gelfand2010}. To avoid most complications from edge effects we imagine in the following that $\ground X$ is finite, and the observation window $W$ is chosen large enough that every point in $Z$ is observed with probability $\approx 1$. Further, we set $w = 1$ for all pairs. Although these simplifying assumptions can often be satisfied in practice, as $\ground X$ is naturally finite and typically entirely observable, smaller ROIs are more convenient to work with, and will then be subject to edge effects. Fortunately, for the size of a typical ROI in SMLM, edge effects should be negligible. 
	
	Choosing the same kernel for both estimators above, an estimator of $\stat{O}{f}$ is
	\begin{equation}
	\estat{O}{f}(r) = \epcf{O}(r)\hat{\gamma}_2^O(f)\emcor{O}{f}(r) = c(r) \sum_{i \neq j}f(t_{o_i},t_{o_j})\kappa(||o_i-o_j|| - r)\mathds{1}_W(o_i,o_j).
	\end{equation}
	
	Rather than computing the mean of $\estat{O}{f}$ directly, we consider the mean of $N^2\estat{O}{f}(r)$, which yields slightly more elegant computations. By splitting the summation according to the cluster and process relationships of pairs, using the symmetry of $f$, and writing $\tilde{c}(r) = N^2c(r)$, we obtain
	\begin{align}
	&\rmean{N^2\estat{O}{f}(r)} \\&= \nonumber \rmean{\tilde{c}(r)\sum_{x\in \ground X}\sum_{(y_1,t_{y_1}), (y_2,t_{y_2}) \in Y_x^2}^{\neq}f(t_{y_1},t_{y_2})\kappa(||y_1-y_2|| - r)\mathds{1}_W(y_1,y_2)} \\&+
	\rmean{\tilde{c}(r)\sum_{ (x_1,x_2) \in \ground X^2 }^{\neq}\sum_{(y_1,t_{y_1}) \in Y_{x_1} }\sum_{(y_2,t_{y_2}) \in Y_{x_2} }f(t_{y_1},t_{y_2})\kappa(||y_1-y_2|| - r)\mathds{1}_W(y_1,y_2)}\nonumber\\&+
	2\rmean{\tilde{c}(r)\sum_{x \in \ground X}\sum_{(y,t_y) \in Y_{x}}\sum_{(e,t_e) \in E}f(t_y,t_e)\kappa(||y-e|| - r)\mathds{1}_W(e,y)}\nonumber\\&+
	\rmean{\tilde{c}(r)\sum_{(e_1,t_{e_1}),(e_2,t_{e_2}) \in E^2}^{\neq}f(t_{e_1},t_{e_2})\kappa(||e_1-e_2|| - r)\mathds{1}_W(e_1,e_2)},\nonumber
	\end{align}
	Using the spatio-temporal dependence structure of our model, we average out the clusters to arrive at
	\begin{align}
	\rmean{N^2\estat{O}{f}(r)} &= 
	\gamma_1(f) n_c \rmean{G} \rmean{ \tilde{c}(r) \sum_{x \in \ground X} (\dispdense*\dispdense)^{\kappa}_x(r)} \\&+
	\gamma_2(f) \rmean{G}^2\rmean{\tilde{c}(r) \sum_{(x_1, x_2) \in \ground X^2}^{\neq} (\dispdense*\dispdense)_{x_1,x_2}^{\kappa}(r)}\nonumber \\&+
	2\gamma_2^{EZ}(f)\rmean{G}\intf{E}\rmean{\tilde{c}(r)\sum_{x \in \ground X^2} (\dispdense*h_e)_x^{\kappa}(r) }\nonumber\\&+
	\gamma_2^E(f)\intf{E}^2\rmean{ \tilde{c}(r) (h_e*h_e)^{\kappa}(r)},\nonumber
	\end{align}
	where
	\begin{align}
	(\dispdense*\dispdense)^{\kappa}_x(r) 
	&= \int \kappa(||t_1-t_2||-r)\dispdense(t_1)\dispdense(t_2)\mathds{1}_{W}(x+t_1,x+t_2)dt_1dt_2, \\
	(\dispdense*\dispdense)_{x_1,x_2}^{\kappa}(r)
	&= \int \kappa(||t_1+x_1-t_2-x_2||-r)\dispdense(t_1)\dispdense(t_2)\mathds{1}_{W}(x_1+t_1,x_2+t_2)dt_1dt_2, \\
	(\dispdense*h_e)_{x}^{\kappa}(r)
	&= \int \kappa(||x+t_1-t_2||-r)\dispdense(t_1)\mathds{1}_{W}(x+t_1,t_2)dt_1dt_2, \\
	(h_e*h_e)^{\kappa}(r)
	&= \int \kappa(||t_1-t_2||-r)\mathds{1}_{W}(t_1,t_2)dt_1dt_2.
	\end{align}
	By considering what happens for $f = 1$ (in which case $\gamma_1(f) = \gamma_2(f) = \gamma_2^E(f) = \gamma_2^{EZ}(f) = 1$), we see that we can rewrite the above as
	\begin{align}
	\rmean{N^2\estat{O}{f}(r)} &= 
	(\gamma_1(f)-\gamma_2(f)) n_c \rmean{G} \rmean{ \tilde{c}(r) \sum_{x \in \ground X} (\dispdense*\dispdense)^{\kappa}_x(r)} \\&+
	\gamma_2(f)\rmean{N^2\epcf{O}(r)}\nonumber\\&+
	2(\gamma_2^{EZ}(f)-\gamma_2(f))\rmean{G}\intf{E}\rmean{\tilde{c}(r)\sum_{x \in \ground X} (\dispdense*h_e)_{x}^{\kappa}(r) }\nonumber\\&+
	(\gamma_2^E(f)-\gamma_2(f))\intf{E}^2\rmean{ \tilde{c}(r) (h_e*h_e)^{\kappa}(r)},\nonumber
	\end{align}
	and we already have a very similar expression to the motion-invariant case. The obstacle to further exact computations come from edge and kernel biases. For the pure cluster term, since we assumed that $Z$ is contained in $W$ with large probability, we have
	\begin{align}
	(\dispdense*\dispdense)^{\kappa}_x(r) &\approx 
	\int \kappa(||t_1-t_2||-r)\dispdense(t_1)\dispdense(t_2)dt_1dt_2  \\
	&=\int \kappa(||t_1||-r)\dispdense(t_1+t_2)dt_1 \dispdense(t_2) dt_2 \\
	&=\int l\kappa(l-r)\dispdense(l\left[cos(\theta),sin(\theta)\right]+t_2)dl d\theta \dispdense(t_2) dt_2 \\
	&= 2\pi \int l\kappa(l-r)(\dispdense*\dispdense)(l)dl,
	\end{align}
	obtained by polar integration, which is a kernel-smoothed version of the cluster autoconvolution. In particular, for small kernel bandwidths, we have 
	\begin{align}
	\rmean{G} \rmean{ \tilde{c}(r) \sum_{x \in \ground X} (\dispdense*\dispdense)^{\kappa}_x(r)} &= \rmean{G}(2\pi r)^{-1}|W|\Lambda_X(W)(\dispdense*\dispdense)^{\kappa}_x(r) \\&\approx |W|\mint{Z}(W) (\dispdense*\dispdense)(r),
	\end{align}
	since $\Lambda_Z(W) \approx \Lambda_X(W) \rmean{G}$ as $Z$ is contained in $W$ with large probability. Using the same tricks for the mixed term, we have 
	\begin{align}
	(\dispdense*h_e)_{x}^{\kappa}(r) &\approx 
	\int \kappa(||x+t_1-t_2||-r)\dispdense(t_1)dt_1dt_2 \\
	&= \int \kappa(||t_2||-r)dt_2 \int\dispdense(t_1)dt_1,
	\end{align}
	so that, for small kernel bandwidths we have
	\begin{align}
	\rmean{G}\rmean{\tilde{c}(r)\sum_{x \in \ground X} (\dispdense*h_e)_{x}^{\kappa}(r) } \approx
	|W|\mint{Z}(W).
	\end{align}
	Finally, for the pure noise term, note that
	\begin{equation}
	\intf{E}^2\rmean{\tilde{c}(r) (h_e*h_e)_{x}^{w,\kappa}(r)} = \rmean{|E\cap W|^2\epcf{E}(r)},
	\end{equation}
	where $\epcf{E}(r)$ is the estimator of the pair correlation function of a stationary Poisson process, so that we can reasonably expect
	\begin{equation}
	\intf{E}^2\rmean{\tilde{c}(r) (h_e*h_e)_{x}^{w,\kappa}(r)} \approx  \rmean{|E\cap W|^2}.
	\end{equation}
	
	Thus, assuming the kernel bandwidth is not too large, we obtain
	\begin{align}
	\rmean{N^2\estat{O}{f}(r)} &\approx
	(\gamma_1(f)-\gamma_2(f)) n_c |W|\mint{Z}(W) (\dispdense*\dispdense)(r) \\&+
	\gamma_2(f)\rmean{N^2\epcf{O}(r)}\nonumber\\&+
	2(\gamma_2^{EZ}(f)-\gamma_2(f))\mint{E}(W)\mint{Z}(W)\nonumber\\&+
	(\gamma_2^E(f)-\gamma_2(f))\rmean{|E\cap W|^2},\nonumber
	\end{align}
	Using simple Taylor expansions for the mean values, we have
	\begin{align}
	\rmean{\estat{O}{f}(r)} &\approx
	(\gamma_1(f)-\gamma_2(f)) n_c \frac{\eta(W)}{\mint{O}(W)|W|^{-1}} (\dispdense*\dispdense)(r) \\&+
	\gamma_2(f)\rmean{\epcf{O}(r)}\nonumber\\&+
	2(\gamma_2^{EZ}(f)-\gamma_2(f))\eta(W)(1-\eta(W))\nonumber\\&+
	(\gamma_2^E(f)-\gamma_2(f))(1-\eta(W))^2,\nonumber
	\end{align}
	or
	\begin{equation}
	\rmean{\estat{O}{f}(r)} \approx (\gamma_1(f)-\gamma_2(f)) \frac{n_c\eta(W)}{\mint{O}(W)|W|^{-1}} (\dispdense*\dispdense)(r) 
	+
	\gamma_2(f)(\rmean{\epcf{O}(r)}-1) + \gamma_2^O(f, W),
	\end{equation}
	where 
	\begin{align}
	\eta(W) &= \frac{\mint{Z}(W)}{\mint{O}(W)}, \\
	\gamma_2^O(f,W) &= \eta(W)^2\gamma_2(f)+(1-\eta(W))^2\gamma_2^E(f)+2\eta(W)(1-\eta(W))\gamma_2^{EZ}(f).
	\end{align}
	Thus, whether $X$ is motion-invariant or not, the mean of the involved summary statistics take approximately the same shape. Note that, since a general distribution for $\ground X$ does not change the spatio-temporal dependence structures, and since all intensity estimation is done for the entire ROI (e.g. the local intensity of $O$ is not needed, only $\Lambda_O(W)$), each term above is naturally estimated by the exact same procedures we developed for the motion-invariant case - the only difference is in interpretation, which must now be conditional on $W$.

\end{document}